\documentclass[12]{article}
\usepackage{amssymb}
\usepackage{a4}
\usepackage{graphicx}
\evensidemargin \oddsidemargin
\newcommand{\beq}{\begin{equation}}
\newcommand{\eeq}{\end{equation}}
\newcommand{\ba}{\begin{array}}
\newcommand{\ea}{\end{array}}
\newcommand{\bea}{\begin{eqnarray}}
\newcommand{\eea}{\end{eqnarray}}

\newcommand{\sect}[1]{\setcounter{equation}{0}\section{#1}}

\newtheorem{lemma}{Lemma}[section]
\newtheorem{theorem}{Theorem}[section]

\newtheorem{naming}{Def.}[section]   
\newcommand{\bdefi}{\medskip\begin{naming} ~ \it}
\newcommand{\edefi}{\end{naming} }
\def\sq{\mbox{\rlap{$\sqcap$}$\sqcup$}}
\newenvironment{proof}[1]{\vspace{5pt}\noindent{\sc Proof #1}\hspace{6pt}}%
{\hfill\sq}
\newcommand{\bp}{\begin{proof}}
\newcommand{\ep}{\end{proof}\par\vspace{10pt}\noindent}
\def\b#1{{\mathbb #1}}

\def\nn{\nonumber  \\}

\date{}

\begin{document}

\title{Existence, uniqueness and stability
for a class of third order dissipative problems depending on time}
 \author{ {\sc  Armando D'Anna$^1$   \hspace{30mm} Gaetano Fiore$^{1,2}$}  \\\\
$^{1}$ Dip. di Matematica e Applicazioni, Universit\`a ``Federico II''\\
   V. Claudio 21, 80125 Napoli, Italy\\         
$^{2}$         I.N.F.N., Sez. di Napoli,
        Complesso MSA, V. Cintia, 80126 Napoli, Italy
       }
 \maketitle

\begin{abstract}
We prove new results regarding the existence, uniqueness, (eventual)
boundedness, (total) stability and attractivity of the solutions of a class of
initial-boundary-value problems characterized by a quasi-linear
third order equation which may contain time-dependent coefficients.
The class includes equations arising in Superconductor Theory and
in the Theory of Viscoelastic Materials. In the proof we use a
Liapunov functional $V$ depending on two parameters,
which we adapt to the characteristics of the problem.
\end{abstract}






\sect{Introduction}

As known, dealing with (in)stability in non-autonomous problems
in general requires careful generalizations of criteria and methods
valid for autonomous problems, even in  linear, finite-dimensional systems
(see e.g. \cite{Ces,LasLef61,Dan82,Mer97,Rio09}). Liapunov direct method 
in its general formulation applies to non-autonomous (as well as to autonomous) 
systems, but the construction of Liapunov functions is more complicated.

In this paper we consider a class of non-autonomous initial-boundary-value problems 
having a number of different physical  applications and 
 prove new results regarding the existence,
uniqueness, boundedness, stability and attractivity of their solutions;
the problems have the form
\bea
\label{eq} && \left\{\ba{l} L\varphi=h(x,t,\Phi), \qquad\quad
L(t):=\partial_t^2\!+\!a \partial_t\!-\!C(t)\partial_x^2
\!-\!\varepsilon(t)
\partial_x^2\partial_t\qquad \qquad
 x\in]0,\!\pi[,\quad t\!>\!t_0, \\[8pt]
\varphi(0,t)=\phi_0(t), \qquad \varphi(\pi,t)=\phi_\pi(t),
\ea\right.\qquad\qquad  \\[8pt]
&& \quad\:\, \varphi(x,t_0)= \varphi_0(x), \qquad \varphi_t(x,t_0)=
\varphi_1(x).\label{eq2}
 \eea
Here $\Phi:=(\varphi,\varphi_x,
\varphi_t)$,
\ $t_0\ge 0$, $\varepsilon\!\in\! C^2(I,I)$,
$C\!\in\! C^1(I,\b{R}^+)$ (with $I\!:=\![0,\infty[$) are functions of $t$, \
with $C(t)\!\ge\! \overline{C}\!=\!\mbox{const}\!>\!0$; \ $a =\mbox{const}$, \
$\varepsilon(t)\!\ge\! 0$, \ $h\in C([0,\pi]\!\times\! I\!\times\!
\b{R}^3)$; $\phi_0,\phi_\pi\in C^2(I)$, $u_0,u_1\in C^2([0,\pi])$
are assigned and fulfill the consistency conditions
\beq
\phi_0(t_0)=\varphi_0(0), \quad \dot \phi_0(t_0)=\varphi_1(0),
\qquad \phi_\pi(t_0)=\varphi_0(\pi), \quad \dot
\phi_\pi(t_0)=\varphi_1(\pi) .  \label{23bis} 
\eeq 
We wish to compare problem (\ref{eq}+\ref{eq2}) to the perturbed one 
\bea \label{eqp}
&& \left\{\ba{ll} Lw=h(x,t,W)+k(x,t),
\qquad \quad  & x\in]0,\!\pi[,\quad t\!>\!t_0, \\[8pt]
w(0,t)=\phi_0(t)\!+\!{\rm w}_0(t), \qquad\qquad & w(\pi,t)=\phi_\pi(t)\!+\!
{\rm w}_\pi(t),
\ea\right.\qquad\qquad  \\[8pt]
&& \quad\:\, w(x,t_0)= \varphi_0(x)\!+\!w_0(x), \qquad w_t(x,t_0)=
\varphi_1(x)\!+\!w_1(x).\label{eq2p}
 \eea
where \ $W:=(w,w_x,
w_t)$, \ $k\in C([0,\pi]\!\times\! I)$, \
 ${\rm w}_0,{\rm w}_\pi\in C^2(I)$, $w_0,w_1\in C^2([0,\pi])$
are assigned and fulfill the consistency conditions
\beq
{\rm w}_0(t_0)=w_0(0), \quad \dot {\rm w}_0(t_0)=w_1(0), \qquad
{\rm w}_\pi(t_0)=w_0(\pi), \quad \dot {\rm w}_\pi(t_0)=w_1(\pi) .
 \label{23bisp}
\eeq
Defining
\beq
\ba{l}
p(x,t):=\frac x{\pi}{\rm w}_\pi(t)+
\left(1\!-\!\frac x{\pi}\right){\rm w}_0(t),
\qquad\qquad u:=w-\varphi-p, \qquad \qquad
u_0(x):=w_0(x)\!-\!p(x,t_0), \\[6pt]
u_1(x):=w_1(x)\!-\!(\partial_tp)(x,t_0) \qquad \quad
f(x,t,U):=h(x,t,U\!+\!\Phi \!+\!P)\!-\!h(x,t,\Phi )\!-\!(L p)(x,t)\!+\!k(x,t),
\ea                                                             \label{deff}
\eeq
where $U:=(u,u_x,
u_t)$, $P:=(p,p_x,
p_t)$,
we find that $u$ fulfills the initial-boundary-value problem
\bea
\label{eqd} && \left\{\ba{l} Lu=f(x,t,U), \qquad \qquad
 x\in]0,\!\pi[,\quad t\!>\!t_0, \label{eq2u}\\[8pt]
u(0,t)\equiv 0, \qquad\qquad u(\pi,t)\equiv 0,  \ea\right.\qquad\qquad  \\[8pt]
&& \quad\:\, u(x,t_0)= u_0(x), \qquad u_t(x,t_0)= u_1(x).\label{eq2d} \eea
$u_0,u_1$ fulfill automatically the consistency
condition $u_0(0)=u_1(0)=u_0(\pi)=u_1(\pi)=0$.
This shows that we can reduce the questions of stability, attractivity of
some $\varphi$ and of boundedness of $w-\varphi$ to those of the
the corresponding $u$ around the origin $u\equiv 0$.
Note that if ${\rm w}_0\equiv {\rm w}_\pi\equiv 0$, then $p\equiv 0$,
$P\equiv 0$, $k\equiv 0$, $f(x,t,0)=0$, and problem (\ref{eq2u}) admits
the null solution, $u(x,t)\equiv 0$.
In (\ref{eq}), (\ref{eqd}) the $\varepsilon$-term is dissipative at t
if $\varepsilon(t)>0$,  and the $a $-term as well if $ a  \!>\!0$.

\begin{figure}[ht]
\begin{center}
\includegraphics[width=6cm]{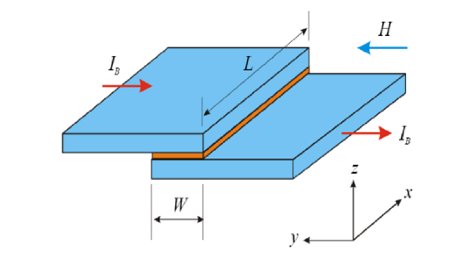}\hfill
\includegraphics[width=3cm]{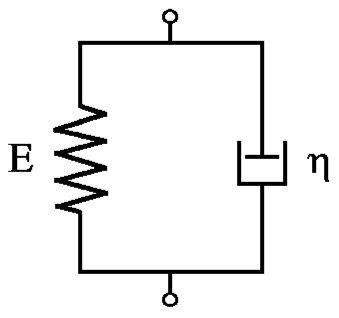}
\end{center}
\caption{Josephson Junction (left) and schematic representation of a Voigt material (right). 
$W,L$ are the width and length of the JJ, \ \ $I_s,H$ are the total 
superconducting carrent  and
the external magnetic field.}
\label{models}       
\end{figure}

Physically remarkable examples of problems (\ref{eq}+\ref{eq2}) include:
\begin{itemize}

\item If $ h\!=\!b\sin \varphi\!-\!\gamma$, with $b,\gamma\!=\!\mbox{const}$,
a modified sine-Gordon  eq. describing
{\bf Josephson effect}  \cite{Jos} in the Theory of Superconductors, which
is at the base (see e.g. \cite{BarPat82}) of a large number
of advanced developments  both in fundamental research (e.g.
macroscopic effects of quantum physics, quantum computation) and in
applications to electronic devices (see e.g. Chapters 3-6 in
\cite{ChrScoSoe99}): $\varphi(x,t)$ is the phase difference of the
macroscopic wavefunctions of the Bose-Einstein condensate of Cooper  pairs
in two  superconductors separated by a   {\it Josephson
junction} (JJ), i.e. a very thin and  narrow   dielectric strip
of finite length (Fig. \ref{models}-left), the $\gamma$ term is the (external) ``bias current''
providing energy to the system,
the term  $ a  \varphi_t$ is due to the Joule effect of the  residual
current of single electrons across the JJ, the term
$\varepsilon \varphi_{xxt}$ is due to the surface impedence of the JJ.
In the simplest  model adopted to describe the
JJ the parameters \ $\varepsilon,C$ \ are constant ($\varepsilon$ is rather small),
and \ $ a  =0$; \ more   accurately,
$ a $ is positive but very small; even  more  accurately,
$ h\!=\!b\sin \varphi\!-\!\gamma-\beta \varphi_t \cos \varphi$ and $\varepsilon,C,\beta$ are positive
($\varepsilon,\beta$ are very small), depend on the
temperature and on the voltage applied to the JJ (see e.g. \cite{LonSco77}),  which
can be controlled and varied with $t$. Also $\gamma$ can be varied with $t$.
Finally, if $\gamma$, or the temperature \cite{KraOboPed97}, or the width of the junction 
\cite{BenCapSco96,PagNapCriEspFruParPelPepSDU95} are spatially dependent, then
new terms linear in $\varphi_x$ may appear in the
equation; in particular
if the width is exponentially shaped the system may be
modelled by the choice $ h\!=\!b\sin \varphi\!-\!\gamma\!-\!\beta \varphi_t \cos
\varphi\!-\!\lambda\varphi_x$ ($\lambda=$const).

\item If \ $ a  \!=\!0$, \ $h=h(x,t)$, \ an equation (see e.g.
\cite{Mor56,Ren83}) for the displacement $\varphi(x,t)$
of the section of a rod from its rest position $x$
in a Voigt material: \ $h$ is applied density force,
$C\!\equiv\! c^2\!=\!E/\rho$, $\varepsilon\!=\!1/\rho\eta$, where $\rho$ is the linear
density of the rod at rest, $E,\eta$ are the elastic and
viscous constants of the rod, which enter the stress-strain relation
$\sigma=E\nu+\partial_t \nu/\eta$, where $\sigma$ is the stress,
$\nu$ is the strain
(as known, a discretized model of the rod is a series
of elements consisting of a viscous damper and an elastic spring connected in parallel
as shown in Fig. \ref{models}-right).
Again, $E,\eta$ may depend on the temperature of
the rod, which can be controlled and varied with $t$.

\item  Equations used to describe: heat conduction at low temperature $\varphi$
\cite{MorPayStr90,FlaRio96,Str11}, if  \ $\varepsilon \!=\!c^2$,  $h=0$; sound
propagation in viscous gases \cite{Lam32}; propagation of plane
waves in perfect incompressible and electrically conducting fluids
\cite{Nar53}.

\end{itemize}

We see that the class of $f$ arising from these examples and (\ref{deff})
is rather broad. 

The plan of the paper is as follows.
As a preliminary step, we prove  (section \ref{existunique}) 
under rather general conditions
existence and uniqueness of solutions of the present problem 
by transforming the latter into one with constant coefficients, for which
known existence and uniqueness theorems \cite{DeaFio12,DanDeaFio12} 
can be applied.
In section \ref{bsa} we present for the non-autonomous problem
(\ref{eq}+\ref{eq2}) boundedness and (asymptotic) stability theorems 
with respect to (w.r.t.) a suitable metric depending 
on $t$ through $\varepsilon(t)$. These  results
generalize those of \cite{DanFio05} - valid for constant
$C,\varepsilon$ - and of \cite{DanFio09,DanFio10} - where we sticked to fixed boundary conditions
and to functions $f$ of the form \ $ f(x,t,U)\!=\! F(u)\!-\!\hat a(x,t,U)u_t$ \
with \ $\hat a\ge 0$\footnote{The latter problem has been recently treated also
in \cite{Rio12UMI}, by an explicit Fourier decomposition of the solution
of problem (\ref{eqd}+\ref{eq2d}).}.
We adopt Liapunov direct method with 
a functional $V$ explicitly depending not only on $t$,
but also on two constant parameters $\gamma,\theta$ which 
we adapt to the characteristics of the problem; their choice and the main properties 
of $V$ are exposed in Lemma \ref{lemma1}.
We then use the powerful comparison method \cite{Yos66}
to study the properties of the solutions of the first order (ordinary) differential inequality
and of the associated  differential equation arising from that lemma; the results are
essential for proving the boundedness, attractivity and stability theorems
for the problem (\ref{eq2u}-\ref{eq2d}).
Moreover, in section \ref{totalstability} we prove for the first time
a  theorem of stability of the solution $u\equiv 0$ of
problem (\ref{eqd}+\ref{eq2d}) when one perturbs also the  right-hand side (rhs)  of 
(\ref{eqd})$_1$ (total stability); in subsection \ref{appl}
we present  two significant classes of forcing terms
to which such a theorem can be applied to study
stability w.r.t. initial, boundary conditions and forcing term perturbations. 
Finally, in section \ref{examples} we illustrate by some examples that one can choose 
coefficients $\varepsilon(t),C(t)$ fulfilling the conditions of the above theorems
and displying rather different behaviours - including periodic ones, or with
$\varepsilon\stackrel{t\!\to\!\infty}{\longrightarrow}0$, or with
$\varepsilon,C\stackrel{t\!\to\!\infty}{\longrightarrow}\infty$.

Throughout the paper we shall use the notation \ $\overline{h}\!:=\!\inf h(t)$, \
$\overline{\overline{h}}\!:=\!\sup h(t)$ \ for any function $h(t)$
defined on $I$.

\sect{Existence and uniqueness of the solution}
\label{existunique}

We are going to prove the existence and uniqueness of the solution
to the above problem under the assumption that \  $\varepsilon\!>\!0$ \ for all $t$.
To this end we note that problem (\ref{eq}+\ref{eq2}) without loss of generality
can be reduced  in two steps 
to one of the same kind where $C,\varepsilon$ are constant.

Indeed, by the change of time variable 
\beq 
t\to\tau(t)\!:=\! \frac 1{ \epsilon }\int\limits_0^t\! \varepsilon(z) dz,
\qquad\quad\Rightarrow\qquad\quad \dot
\tau=\frac {\varepsilon}{\epsilon}, \qquad
\partial_t=\frac {\varepsilon}{\epsilon}\partial_\tau,
\qquad\partial_t^2=\frac{\varepsilon^2}{\epsilon^2}\partial_\tau^2+\frac
{\dot\varepsilon}{\epsilon}\partial_\tau
\eeq
(here $\epsilon$ is a positive constant with the same dimensions as $\varepsilon$)
we transform (\ref{eq}) into a new equation of the form
$(\varphi_{\tau\tau}-\epsilon\varphi_{xx\tau})\varepsilon^2[t(\tau)]/\epsilon^2=...$,
where the dots stand for an expression not depending on 
$\varphi_{\tau\tau},\varphi_{xx\tau}$. By the change of the dependent
variable
\beq
 \varphi\to
\tilde\varphi(x,\tau)\!:=\!b^{-1}(\tau)\:\varphi[x,t(\tau)], \qquad
\qquad b(\tau):= \exp\!\int^\tau_0\!\!\!dz\left\{\frac
1{\epsilon}-\frac {\epsilon C[t(z)] }{\varepsilon^2[t(z)] }\right\} 
\eeq
the equation is further transformed into $(\tilde\varphi_{\tau\tau}\!-\!
\tilde\varphi_{xx}-
\epsilon\tilde\varphi_{xx\tau})b\varepsilon^2/\epsilon^2=...$,
where the dots stand for an expression not depending on 
$\tilde\varphi_{\tau\tau},\tilde\varphi_{xx},\tilde\varphi_{xx\tau}$. 
Multiplying by $\epsilon^2/b\varepsilon^2$ 
(by the above assumptions, in any finite interval $[0,T]$
 this function is bounded from above by a finite constant)
the problem (\ref{eq}) \& (\ref{eq2}) takes the final equivalent form
\bea \label{eqg'}
&& \left\{\ba{l} \tilde\varphi_{\tau\tau}\!-\! \tilde\varphi_{xx}\!-\! \epsilon
\tilde\varphi_{xx\tau}=\tilde f(x,\tau,\tilde\varphi,\tilde\varphi_x,
\tilde\varphi_\tau) ,\qquad\quad
 x\in]0,\!\pi[,\qquad  T\!>\! \tau\!>\!\tau_0, \\[8pt]
\tilde\varphi(0,\tau)=\tilde \phi_0(\tau), \quad \tilde\varphi(\pi,\tau)=\tilde
\phi_\pi(\tau),  \ea\right.\qquad\qquad  \\[8pt]
&& \quad\:\, \tilde\varphi(x,\tau_0)=\tilde\varphi_0(x) , \qquad
\tilde\varphi_\tau(x,\tau_0)=\tilde\varphi_1(x) ,\label{eq2g'}
 \eea
where
\bea
&&\tau_0\!:=\! \frac 1{\epsilon }\int\limits_0^{t_0}\! \varepsilon(z) dz,
\qquad\tilde\varphi_0(x) \!:=\! \frac{\varphi_0(x)}{b(\tau_0)}, \quad
\tilde\varphi_1(x)\!:=\!
\frac{\epsilon\varphi_1(x)}{b(\tau_0)\varepsilon(t_0)}
\!-\!\frac{b_\tau(\tau_0)}{b^2(\tau_0)}\varphi_0(x),
\qquad
\tilde \phi_i(\tau)\!:=\!\frac{ \phi_i[t(\tau)]}{b(\tau)}, \quad i=0,\pi,\nn
&&\tilde f(x,\tau,\tilde\varphi,\tilde\varphi_x,
\tilde\varphi_\tau):=
\left[\frac {\epsilon^2}{b\varepsilon^2}
h\!\left(\!x,t,b\tilde\varphi,b\tilde\varphi_x,
\frac {b\varepsilon}{\epsilon } \tilde\varphi_\tau\!+\!\frac {b_\tau\varepsilon}{\epsilon } \tilde\varphi
\!\right)-\left(\!a \frac {\varepsilon}{\epsilon }\!+\!2\frac {b_\tau}b\!+\!
\frac {\dot\varepsilon\epsilon }{\varepsilon^2}\!\right) \tilde\varphi_\tau
+\left(\!\frac {b_{\tau\tau}}b\!+\!\frac {a  b_\tau\epsilon}{b\varepsilon}
\!+\!\frac { b_\tau\epsilon\dot\varepsilon}{b\varepsilon^2}
\!\right)\tilde\varphi\right]_{t=t(\tau)}.
\nonumber
\eea
Under the above assumptions
on $C,\varepsilon$, \ $\tilde f$ \ is locally Lipschitz iff $f$ is.
This is now a problem already treated in \cite{
DeaFio12},
where a theorem of existence and uniqueness of the solution has been proved
formulating it as an equivalent integral-differential
equation in any time interval $[0,T]$, and applying the fixed-point theorem.
This theorem can be applied now to the present context.
[Note that requiring  \ $\varepsilon\!>\!0$ \ for all $t$ we  are not excluding that
$\varepsilon$ may go to zero as \ $t\to\infty$: in any finite $[0,T]$ it
is in any case $\inf_{[0,T]} \varepsilon(t)\!>\!0$, so that the above definitions are
safe.]
 We hope to further generalize this theorem elsewhere.

\sect{Boundedness, stability, attractivity}
\label{bsa}

\subsection{Preliminaries}
\label{preliminaries}

The  solution $\varphi$ of problem (\ref{eq}+\ref{eq2})
and the solution $w$ of the perturbed problem (\ref{eqp}+\ref{eq2p})
are `close' to each other iff $u$
is a `small' solution of problem (\ref{eqd}+\ref{eq2d}) and coincide iff $u$
is the null solution. We give a precise meaning to this
introducing the distance between $\varphi,w$ as the  norm
$d(u,u_t)$ of $u$, where the $t$-dependent norm $d(\varphi,\psi)$ is defined by
\beq
d^2(\varphi,\psi)\equiv
d_\varepsilon^2(\varphi,\psi)=\int\limits_0^\pi\!\!dx\,[\varepsilon^2(t)
\varphi_{xx}^2\!+\!\varphi_x^2\!+\!\varphi^2\!+\!\psi^2];
\eeq
 $\varepsilon^2$ plays the role of a weight for the
second order derivative term $\varphi_{xx}^2$ so that
for $\varepsilon=0$ this automatically reduces to the proper norm needed
for treating the corresponding second order problem. Using the condition
$\varphi(0)=0=\varphi(\pi)$ one easily derives that
\beq
|\varphi(x)|\le d(\varphi,\psi),\qquad\qquad\varepsilon|\varphi(x)| \le \pi^{\frac 32}
d(\varphi,\psi), \qquad\qquad
\varepsilon|\varphi_x(x)| \le \pi^{\frac 12} d(\varphi,\psi),
 \label{3ineq}
\eeq
for any $x$.\footnote{From $\varphi^2(x)\!=\!\int_0^xdx'\frac {d\varphi^2}{dx}(x')
\!=\!\int_0^xdx'2\varphi(x')\varphi_x(x')$ and $2|\varphi\varphi_x|\!\le\!(\varphi\!+\!
\varphi_x^2)$ it follows $\varphi^2(x)\!\le\!\int_0^xdx'(\varphi\!+\!\varphi_x^2)
 \!\le\!\int_0^\pi dx'(\varphi\!+\!\varphi_x^2)\le d^2(\varphi,\psi)$, as claimed.
From $\varphi(0)=0=\varphi(\pi)$ it follows that there exists $\xi\in]0,\pi[$ such
that $\varphi_x(\xi)=0$; hence, using Schwarz inequality,
$$
\varphi_x(x)=\varphi_x(x)\!-\!\varphi_x(\xi)=\int\limits^x_\xi\!\!\varphi_{yy}(y) dy \quad\Rightarrow
\quad \varepsilon|\varphi_x(x)| \le \left|\!\!\int\limits^x_\xi\!\!|\varepsilon\varphi_{yy}(y)| dy
\right| \le \!\! \int\limits^\pi_0\!\!|\varepsilon\varphi_{yy}(y)| dy
\le\! \left[\! \int\limits^\pi_0\!\!dy\right]^{\frac 12}\!
 \left[\!\int\limits^\pi_0\!\!|\varepsilon\varphi_{yy}(y)|^2 dy\right]^{\frac 12}\!\!\le
 \!\pi^{\frac 12} d
$$
as claimed. Finally, from $\varphi(x)=\int\limits^x_0\!\!\varphi_{y}(y) dy$
it follows $|\varepsilon\varphi(x)|\le\int\limits^x_0\!\!|\varepsilon\varphi_{y}(y)| dy
 \le\int\limits^\pi_0\!\!|\varepsilon\varphi_{y}(y)| dy\le \int\limits^\pi_0\!\!\pi^{\frac 12} d\, dy= \pi^{\frac 32} d$, as claimed.
}
Therefore a convergence in the norm $d$ implies also a
uniform (in $x$) pointwise convergence of  $\varphi$ and a
uniform (in $x$) pointwise convergence of  $\varphi_x$ for
$\varepsilon(t)\!\neq\! 0$.

The notions of (eventual) boundedness,
stability, attractivity, etc.
are formulated using this distance, i.e. the norm of $u$, which
we shall abbreviate as $d(t)\equiv d_{\varepsilon(t)}\big[u(x,t),u_t(x,t)\big]$ whenever this is not ambiguous.
Therefore, without loss of generality,
to investigate these properties we  consider problem
(\ref{eqd}+\ref{eq2d}).

\bdefi \label{def-1} \rm The solutions of (\ref{eq}-\ref{eq2}) are
{\it bounded}  if for any \ $\alpha\!>\!0,\: t_0\!>\!0$
\ there exists \ $\beta(\alpha,t_0)\!>\!0$ \ such that
 $$
 d(u_0,u_1)\leq \alpha\qquad \Rightarrow\qquad
 d(u,u_t)<\beta\:\quad\forall t\ge t_0;
 $$
 {\it eventually bounded}  if $\exists s(\alpha)\!\ge\! 0$ such that
this  holds for $t_0\!\ge\!s$;
{\it (eventually) uniformly bounded} if $\beta=\beta(\alpha)$.
\edefi

\bdefi \label{def-2} \rm \ $u(x,t)\equiv 0$ \ is  {\it eventually quasi-uniform-asymptotically
stable in the large} if there exists a $\bar t\ge 0$
such that  for any \  $\rho,\alpha>0$  \ there exist \  $s(\alpha)\ge \bar t$
\  such that
$$
d(u_0,u_1)<\alpha, \quad t_0\ge s(\alpha)\qquad \quad\Rightarrow\qquad \quad
\exists T(\rho,\alpha,t_0)>0\quad\mbox{s. t. }\quad d(u,u_t)<\rho\:\quad\forall t\ge t_0\!+\!T.
$$
It is {\it quasi-uniform-asymptotically stable in the large} if this holds with $s(\alpha)=\bar t$
and $T=T(\rho,\alpha)>0$.
\edefi

Suppose now that \  $f(x,t,0)=0$, \  so that  \ $u(x,t)\equiv 0$ \ is a solution of
problem (\ref{eqd}+\ref{eq2d}).
\ If $f$ is defined as in (\ref{deff}) this occurs
for $p\equiv 0$, $k\equiv 0$.

\bdefi \label{def-3} \rm $u(x,t)\equiv 0$  is
{\it stable} if for any \  $\sigma>0$ 
\ there exists a \ $\delta(\sigma,t_0)> 0$ \  such that
$$
d(u_0,u_1)<\delta(\sigma,t_0)\qquad \Rightarrow\qquad
d(u,u_t)<\sigma\:\quad\forall t\ge t_0.
$$
It is {\it uniformly stable} if $\delta=\delta(\sigma)$. \edefi

\medskip
We introduce the non-autonomous family of Liapunov functionals
\beq
\label{321}
V\equiv V(\varphi,\psi,t;\gamma,\theta):=\int_0^\pi\!
\frac{1}{2}\!\Big\{\!\gamma\psi^2\!+\!(\varepsilon
\varphi_{xx}\!-\!\psi)^2\!\!+
[C(1\!+\!\gamma\!)\!-\!\dot\varepsilon\!+\!\varepsilon(a \!+\!\theta)]
\varphi_x^2+a \theta\varphi^2\!+\!2\theta\varphi\psi \Big\}dx\nonumber
\eeq
where $\theta,\gamma$ are for the moment  unspecified positive
parameters.
$V$ reduces to the Liapunov functional of
\cite{DanFio05} for constant $\varepsilon$, $C\equiv 1$, $\theta=a =0$.

\subsection{Main assumptions and preliminary estimates}

\begin{figure}[ht]
\begin{center}
\includegraphics[width=5.5cm]{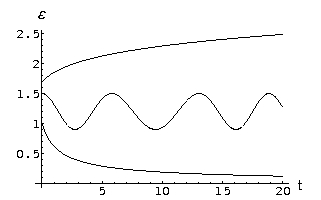}
\hfill 
\includegraphics[width=5.5cm]{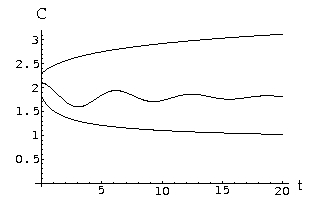}
\end{center}
\caption{Possible $t$-dependences for $\varepsilon,C$}
\label{fig:2}       
\end{figure}

We consider rather general $t$-dependences for $\varepsilon,C$,
like the ones depicted in figures \ref{fig:2}. To be more precise,
 denote
$\dot C_+(t)\!:=\!\left\{\ba{ll} \dot C(t) \quad &\mbox{ if }
\dot C(t)\ge 0\\ 0 \quad &\mbox{ if } \dot C(t)<0,\ea\right.$, \
$\dot C_-\!:=\!\dot C\!\!-\!\!\dot C_+$.
We assume that there exists a constant $\mu\!>\!0$ such that
\beq
\overline{C}\!>\!0,\qquad\quad \overline{\varepsilon}\!\ge\!0,\qquad\quad
a \!+\!\overline{\varepsilon}>0,
\qquad\quad
C\!-\!\dot\varepsilon\!\ge\!\mu(1\!+\!\varepsilon),\qquad\quad
 \mu\varepsilon^2\!+\!\mu \varepsilon \!+\!
\ddot\varepsilon \!-\!(1\!+\!\gamma)\dot C_->0 \label{condi4}
\eeq
We are not excluding the following cases: $\varepsilon(t)=0$ for
some $t$, $\varepsilon\stackrel{t\!\to\!\infty}{\longrightarrow}0$,
$\varepsilon(t)\equiv 0$, 
$\varepsilon\stackrel{t\!\to\!\infty}{\longrightarrow}\infty$ [in
view of (\ref{condi4})$_4$ the latter condition requires also
$C\stackrel{t\!\to\!\infty}{\longrightarrow}\infty$]; but, by
condition (\ref{condi4})$_3$, in (\ref{eq}) either the term containing
$a $ or the one containing $\varepsilon$
in (\ref{eq})  is dissipative.

We recall Poincar\'e inequality, which easily follows from Fourier analysis:
\beq
\phi\in C^1([0,\pi]),\quad\phi(0)=0,
\quad\phi(\pi)=0,\qquad\Rightarrow \qquad \int^\pi_0 \!\!\!
dx\,\phi_x^2(x)\ge \int^\pi_0 \!\!\! dx\,\phi^2(x). \label{poinc}
\eeq

\subsection{\normalsize Upper bound for $V$}

From definition (\ref{321}) and the inequalities
$-2\varepsilon\varphi_{xx}\psi\!\le\!\varepsilon^2\varphi^2_{xx}\!+\!\psi^2$,
$2\theta\varphi\psi\!\le\!\theta(\varphi^2\!+\!\psi^2)$,
(\ref{condi4})$_4$
 we easily find
\bea && V(\varphi,\psi,t;\gamma,\theta)\!\le\!  \int\limits_0^\pi\!
\frac{1}{2}\!\left\{\!(\gamma\!+\!2\!+\!\theta)\psi^2\!+\!2\varepsilon^2
\varphi_{xx}^2\!\!+\!\left[C(1\!+\!\gamma\!)\!-\!\dot\varepsilon
\!+\!\varepsilon(a \!+\!\theta)\right]\varphi_x^2\!+\!(a \!+\!1)\theta\varphi^2
\right\}dx \nn && \qquad
<\!\int\limits_0^\pi\!
\frac{1}{2}\!\left\{\!(\gamma\!+\!2\!+\!\theta)\psi^2\!+\!2\varepsilon^2
\varphi_{xx}^2 \!\!+\!\left[C\gamma\!+\!(C
\!-\!\dot\varepsilon)\!\left(1\!+\!\frac{a \!+\!\theta}{\mu}\right)\!\right]\varphi_x^2
\!+\!(a \!+\!1)\theta\varphi^2\right\}dx \nonumber
\eea
for all $\gamma>0,\theta\ge \theta_0:=\max\{0,-a \}$.
By (\ref{condi4})$_3$ there exists $\lambda\!\in]0,1[$ such that $a \!+\! \lambda\overline{\varepsilon}\!>\!0$.
Choosing
\bea
&& \theta> \theta_1=\max\left\{2a ,
\frac {-a }{1\!-\!\lambda}\right\},\label{defthetatheta1}     \\[14pt] &&
\gamma\ge\gamma_1:=2+\max\left\{\frac{a \!+\!\theta}{\mu},(|a |\!+\!1)\theta,
\frac{4\theta}{\overline{\varepsilon}\lambda\!+\! a },
2\frac {\theta\!-\!a }{\overline{\varepsilon}\!+\! a }\right\}
\label{defthetagamma1}
\eea
and setting
\beq
G(t)\!:=\!C(t)\!-\!\frac 12\dot\varepsilon(t)\!+\! 1 \label{defgB}
\eeq
[note that by (\ref{condi4}) it is $G(t)\!>\!1$] we find that
\bea
V(\varphi,\psi,t;\gamma,\theta)&\!\le\! &\frac{1}{2}
\int\limits_0^\pi\!
\!\left[\!2\gamma\psi^2\!+\!2\gamma\varepsilon^2
\varphi_{xx}^2\!\!+\!\gamma\left(2C
\!-\!\dot\varepsilon\right)\varphi_x^2\!+\!2\gamma\varphi^2\right]dx
\le   \gamma \, G(t)\, d^2.                                 \label{Ineq3}
\eea

\subsection{\normalsize Lower bound for $V$}

 We find, on one hand,
\bea
&&(\varepsilon\varphi_{xx}\!-\!\psi)^2=\frac{\varepsilon^2}2\varphi_{xx}^2
\!+\!\left(\frac{\varepsilon}{\sqrt{2}}\varphi_{xx}
\!-\!\sqrt{2}\psi\right)^2\!-\!\psi^2
\ge \frac{\varepsilon^2}2\varphi_{xx}^2 \!-\!\psi^2,      \label{i1}\\
&&(\varepsilon\lambda\!+\! a )\theta\varphi^2\!+\!2\theta\varphi\psi
= \frac 34 (\varepsilon\lambda\!+\! a )\theta\varphi^2\!+\!
(\varepsilon\lambda\!+\! a )\theta\left[\frac{\varphi}2\!+\!
\frac{2\psi}{\varepsilon\lambda\!+\! a }\right]^2
\!-\!\frac{4\theta}{\varepsilon\lambda\!+\! a }\psi^2, \qquad\quad    \label{i2}\\
&&C(1\!+\!\gamma\!)\!-\!\dot\varepsilon\!+\!\varepsilon\big[a \!+\!(1\!-\!\lambda)
\theta\big]\ge
\mu(1\!+\!\varepsilon)\!+\!C\gamma\!+\! \varepsilon\big(a \!+\!(1\!-\!\lambda)
\theta\big)\ge \mu(1\!+\!\overline{\varepsilon})\!+\!\overline{C}\gamma\!+\!
\overline{\varepsilon}\big[a \!+\!(1\!-\!\lambda)\theta\big]. \qquad\quad   \label{i3}
\eea
In the last line we have used (\ref{condi4}), (\ref{defthetatheta1}). On the other hand,
\beq
\label{321'}
V=\frac{1}{2}\!\int\limits_0^\pi\! \!\Big\{\!\gamma\psi^2\!+\!(\varepsilon
\varphi_{xx}\!-\!\psi)^2\!\!+\!
\left[C(1\!+\!\gamma\!)\!-\!\dot\varepsilon\!+\!
\varepsilon\big(a \!+\!(1\!-\!\lambda)\theta\big)\right]\varphi_x^2\!+\!
\varepsilon\lambda\theta(\varphi_x^2\!-\!\varphi^2)\!+\!
(\varepsilon\lambda\!+\! a )\theta\varphi^2\!+\!2\theta\varphi\psi \Big\}dx.\qquad
\eeq
From (\ref{i1}-\ref{321'}), (\ref{poinc}) and $\overline{\varepsilon}\lambda\!+\!a >0$ we find
\bea
V & \ge & \frac{1}{2}\!\int\limits_0^\pi\!\!\left\{\!\left[\gamma\!-\!1\!-\!
\frac{4\theta}{\varepsilon\lambda\!+\! a }\right]\psi^2\!+\!
\frac{\varepsilon^2}2\varphi_{xx}^2
\!\!+\!\left[\mu(1\!+\!\overline{\varepsilon})\!+\!\overline{C}\gamma
\!+\! \overline{\varepsilon}\big(a \!+\!(1\!-\!\lambda)\theta\big)
\right]\varphi_x^2\!+\!
\frac 34 (\varepsilon\lambda\!+\! a )\theta\varphi^2 \right\}dx\nn
& \ge & \chi_0 d^2, \qquad \quad
\chi_0:=\frac 12\min\left\{\frac 12, \mu
(1\!+\!\overline{\varepsilon})\!+\!\overline{C}\gamma\!+\!
\overline{\varepsilon}\big(a \!+\!(1\!-\!\lambda)\theta\big),
\gamma\!-\!1\!-\!\frac{4\theta}{\overline{\varepsilon}\lambda\!+\! a },
\frac 34 (\varepsilon\lambda\!+\! a )\theta
\right\}.\qquad               \label{LBV}
\eea
$\chi_0$ is positive by (\ref{defthetagamma1}).

\subsection{\normalsize Upper bound for $\dot V$}

Let $V(t;\gamma,\theta)\!:=\!V(u,u_t,t;\gamma,\theta)$.
Reasoning as in Ref. \cite{DanFio09} we find
\bea
&&\dot
V(t;\gamma,\theta)=\int\limits_0^\pi\! \!\left\{\!(\varepsilon
u_{xx}\!-\!u_t)(\varepsilon u_{xxt} \!-\!u_{tt}\!+\!\dot\varepsilon
u_{xx}) \!+\![\dot C(1\!+\!\gamma\!)\!-\!\ddot\varepsilon\!+\!\dot
\varepsilon(a \!+\!\theta)] \frac{u_x^2}2\right.
\nn &&\qquad\qquad
\left.\!+\![C(1\!+\!\gamma\!)\!-\!\dot\varepsilon\!+\!
\varepsilon(a \!+\!\theta)]u_xu_{xt}
\!+\!a \theta uu_t\!+\!\theta u_t^2\!+\!(\gamma u_t\!+\!\theta
u)u_{tt} \right\}dx
\nn &&=\int\limits_0^\pi\! \!\left\{\!(\varepsilon
u_{xx}\!-\!u_t)[a u_t \!-\! Cu_{xx}\!-\!f
\!+\!\dot\varepsilon u_{xx}] \!+\![\dot
C(1\!+\!\gamma\!)\!-\!\ddot\varepsilon\!+\!\dot
\varepsilon(a \!+\!\theta)] \frac{u_x^2}2\right.
\nn &&\qquad
\left.-[C(1\!+\!\gamma\!)\!-\!\dot\varepsilon
\!+\!\varepsilon(a \!+\!\theta)]u_{xx}u_t \!+\!a \theta
uu_t\!+\!\theta u_t^2\!+\!(\gamma
u_t\!+\!\theta u)[Cu_{xx}\!+\!\varepsilon
u_{xxt}\!-\!a u_t\!+\!f] \right\}dx
\nn && = \!\int\limits_0^\pi\! \!\Big\{\!\varepsilon
u_{xx}[(\dot\varepsilon\!\!-\!\!
C)u_{xx}\!\!-\!\!f]\!+\!u_t[a \varepsilon
u_{xx}\!-\!a u_t \!+\! Cu_{xx}\!+\!f
\!-\!\dot\varepsilon u_{xx}\!-\! C(1\!+\!\gamma\!)u_{xx}
\nn &&\qquad
+\dot\varepsilon
u_{xx}\!-\!\varepsilon(a \!\!+\!\theta)u_{xx}
\!+\!\theta u_t\!+\! \gamma Cu_{xx}\!+\!\gamma\varepsilon
u_{xxt}\!+\!\gamma f\!-\!a \gamma u_t]
\nn &&\qquad \left.
+\theta u[Cu_{xx}\!+\!\varepsilon
u_{xxt}\!+\!f] \!+\![\dot C(1\!+\!\gamma\!)\!-\!\ddot\varepsilon\!+\!\dot
\varepsilon(a \!+\!\theta)]\frac{u_x^2}2\right\}dx
\nn &&=
\int\limits_0^\pi\! \!\Big\{\!\varepsilon [(\dot\varepsilon\!-\!
C)u_{xx}\!-\!f]u_{xx}\!+\!u_t[(f\!-\!a u_t)(1\!+\!\gamma\!)
\!-\!\varepsilon\theta u_{xx} \!+\!\theta u_t\!+\!\gamma\varepsilon u_{xxt}]
\nn &&\qquad\quad \left. +\theta u[Cu_{xx}\!+\!\varepsilon
u_{xxt}\!+\!f] \!+\![\dot
C(1\!+\!\gamma\!)\!-\!\ddot\varepsilon\!+\!\dot
\varepsilon(a \!+\!\theta)] \frac{u_x^2}2\right\}dx
\nn && =-\!\int\limits_0^\pi \!\!\left\{\!\varepsilon
[(C\!-\!\dot\varepsilon)
u^2_{xx}\!+\! fu_{xx}]\!+\!
\left[a (1\!+\!\gamma)\!-\!\theta\right]u_t^2\!+\!
\left[2\theta C\!+\!\ddot\varepsilon\!-\!
\dot\varepsilon(a \!+\!\theta)\!-\!(1\!+\!\gamma)\dot
C\right]\frac{u_x^2}2\right.
\nn[8pt] &&
\qquad\qquad \!+\!   \varepsilon\gamma u^2_{xt}  \!-\!
[(1\!+\!\gamma)u_t\!+\!\theta u ]f\Big\}dx.         \label{e1}
\eea
We shall use the following inequalities: (\ref{condi4}), (\ref{poinc}) and
\bea
&&\int\limits_0^\pi \!\!\varepsilon [(C\!-\!\dot\varepsilon)
u^2_{xx}\!+\! fu_{xx}]dx\ge
\int\limits_0^\pi \!\!\varepsilon [\mu(1\!+\!\varepsilon)
u^2_{xx}\!+\! fu_{xx}]dx
 =\int\limits_0^\pi \!\!\left[\varepsilon \mu
u^2_{xx}\!+\!\frac 34 \mu\varepsilon^2u^2_{xx}\!+\!
\left(\frac {\sqrt{\mu}}2 \varepsilon u_{xx}\!+\!
\frac f{\sqrt{\mu}}\right)^2\!\!-\!\frac {f^2}{\mu}\right]\!dx
\nn &&\qquad \ge\int\limits_0^\pi \!\!\left[\varepsilon \mu
u^2_{xx}\!+\!\frac 34 \mu\varepsilon^2u^2_{xx}
\!-\!\frac {f^2}{\mu}\right]\!dx;         \label{d1}\\
&&\int\limits_0^\pi \!\!\theta\!\left[\frac C4 u_x^2\!-\! u f\right]dx
\ge\int\limits_0^\pi \!\!\theta\!\left[\frac {\overline{C}}4
u_x^2 \!-\! u f\right]dx
=\int\limits_0^\pi \!\!\frac {\theta}4\!\left[\overline{C}
(u_x^2\!-\!u^2)\!+\! \left(\!\sqrt{\overline{C}} u\!-\!
\frac {2f}{\sqrt{\overline{C}}}\!\right)^2\!\!\!-\!  \frac {4f^2}
{\overline{C}}\right]\!dx \ge-\!\int\limits_0^\pi \!\!\frac {\theta f^2}
{\overline{C}}dx;      \qquad\qquad                  \label{d2}\\
&&\int\limits_0^\pi \!\!\left\{\!\varepsilon\gamma
u^2_{xt} \!+\!\left[a (1\!+\!\gamma)\!-\!\theta\right]u_t^2 \!-\!
(1\!+\!\gamma)fu_t\right\}dx
= \int\limits_0^\pi \!\!\left\{\!\varepsilon\gamma
(u^2_{xt}\!-\!u_t^2)  \!+\!\left[\varepsilon\gamma\!+\!a 
(1\!+\!\gamma)\!-\!\theta\right]u_t^2 \!-\!
(1\!+\!\gamma)fu_t\right\}dx\nn
&&=\int\limits_0^\pi \!\!\left\{\!\varepsilon\gamma
(u^2_{xt}\!-\!u_t^2)  \!+\!\left[(\varepsilon\!+\!a )\frac{\gamma\!-\!1}2
\!+\!a \!-\!\theta\right]u_t^2 \!+\!\frac{\gamma\!+\!1}2
\left[\sqrt{\varepsilon\!+\!a }u_t\!-\! \frac f{\sqrt{\varepsilon\!+\!a }}
\right]^2\!-\! \frac{\gamma\!+\!1}{\varepsilon\!+\!a }\frac {f^2}2
\right\}dx\nn
&&\qquad \ge \int\limits_0^\pi \!\!\left\{\!\left[(\varepsilon\!+\!a )\frac{\gamma\!-\!1}2 \!+\!a \!-\!\theta\right]u_t^2 \!-\!
\frac{\gamma\!+\!1}{\varepsilon\!+\!a }\frac {f^2}2
\right\}dx;                          \label{d3}\\
&&\int\limits_0^\pi\!\!\left[\frac 32\theta C\!+\!\ddot\varepsilon\!-\!
\dot\varepsilon(a \!+\!\theta)\!-\!(1\!+\!\gamma)\dot
C\right]\frac{u_x^2}2dx =\int\limits_0^\pi\!\!\left[
C\left(\frac {\theta}2\!-\!a \right)\!+\!\ddot\varepsilon \!+\!(C\!-\!
\dot\varepsilon)(a \!+\!\theta)\!-\!(1\!+\!\gamma)\dot
C\right]\frac{u_x^2}2dx\nn
&&\qquad \ge\int\limits_0^\pi\!\!\left[
\overline{C}\left(\frac {\theta}2\!-\!a \right)\!+\!\ddot\varepsilon \!+\!
\mu(1\!+\!\varepsilon)(a \!+\!\theta)\!-\!(1\!+\!\gamma)(\dot
C_-\!+\! \dot C_+)\right]\frac{u_x^2}2dx.              \label{d4}
\eea
From (\ref{e1}-\ref{d4}) we obtain
\bea
&&\dot V\le-\!\int\limits_0^\pi \!\!\left\{\!
\varepsilon \mu u^2_{xx}\!+\!\frac 34 \mu\varepsilon^2u^2_{xx}
\!-\!\frac {f^2}{\mu}\!+\!
\left[(\varepsilon\!+\!a )\frac{\gamma\!-\!1}2 \!+\!a \!-\!
\theta\right]u_t^2 \!-\! \frac{\gamma\!+\!1}{\varepsilon\!+\!a }\frac {f^2}2
\right.\nn
&&\qquad \qquad\!+\!\left.\left[\overline{C}\left(\frac {\theta}2\!-\!a \right)
\!+\!\ddot\varepsilon \!+\!\mu(1\!+\!\varepsilon)(a \!+\!\theta)
\!-\!(1\!+\!\gamma)(\dot C_-\!+\! \dot C_+)\right]\frac{u_x^2}2
 \!-\! \frac {f^2\theta} {\overline{C}}\right\}dx \nn
&&=-\!\int\limits_0^\pi \!\!\left\{\! \frac
{\mu}8\varepsilon^2u^2_{xx} \!+\!\frac {5\mu}8
\varepsilon^2(u^2_{xx}\!-\!u^2_x) \!+\!\frac {\mu}8
\varepsilon^2u^2_x\!+\!\frac {\mu}2 \varepsilon^2u^2_x\!+\!\frac
{\mu}2 \varepsilon u^2_{xx}\!+\!\frac {\mu}2 \varepsilon
(u^2_{xx}\!-\!u^2_x) \!+\!\frac {\mu}2 \varepsilon u^2_x\right.
\nn &&\qquad\!+\!\left.
\left[(\varepsilon\!+\!a )\frac{\gamma\!-\!1}2
\!+\!a \!-\!\theta\right]u_t^2
\!+\!\left[\ddot\varepsilon \!+\!\mu\varepsilon(a \!+\!\theta)
\!-\!(1\!+\!\gamma)\dot C_- \!+\!\overline{C}\left(\!\frac
{\theta}2\!-\!a \!\right)
\!+\!\mu(a \!+\!\theta)\right]\frac{u_x^2}2\right\}dx
\nn &&\qquad\quad
\!+\!\int\limits_0^\pi\!\!\left[\frac 1{\mu}\!+\!
\frac{\gamma\!+\!1}{2(\varepsilon\!+\!a )} \!+\! \frac {\theta}
{\overline{C}}\right]f^2dx +\frac{1\!+\!\gamma}2\dot
C_+\int\limits_0^\pi\!\!\frac{u_x^2}2dx\nn
&&\le-\!\int\limits_0^\pi
\!\!\left\{\! \frac {\mu}8\varepsilon^2u^2_{xx}\!+\!\frac
{\mu}2 \varepsilon u^2_{xx} \!+\!
\left[(\varepsilon\!+\!a )\frac{\gamma\!-\!1}2
\!+\!a \!-\!\theta\right]u_t^2\!+\!\left[\frac {\mu}4
\overline{\varepsilon}^2 \!+\!\overline{C}\left(\!\frac
{\theta}2\!-\!a \!\right) \!+\!\mu(a \!+\!\theta)\right]\frac{u_x^2}2
+\right.\nn &&\left.
 \left[\mu\varepsilon^2\!+\!\mu \varepsilon \!+\!
\ddot\varepsilon \!+\!\mu\varepsilon(a \!\!+\!\theta)
\!-\!(1\!\!+\!\!\gamma)\dot C_-\right]\frac{u_x^2}2\right\}dx
\!+\!\int\limits_0^\pi\!\!\left[\frac 1{\mu}\!+\!
\frac{\gamma\!+\!1}{2(\varepsilon\!\!+\!\!a )} \!+\! \frac {\theta}
{\overline{C}}\right]\!f^2dx \!+\!\frac{1\!\!+\!\!\gamma}2\dot
C_+\!\!\int\limits_0^\pi\!\!\frac{u_x^2}2dx
\eea
By the choices (\ref{defthetatheta1}-\ref{defthetagamma1}) of $\theta,\gamma$
we find
$$
(\overline{\varepsilon}\!+\!a )\frac{\gamma\!-\!1}2 \!+\!a \!-\!\theta
\:\ge\: (\overline{\varepsilon}\!+\!a )\left[2\!+\!2\frac {\theta\!-\!a }
{\overline{\varepsilon}\!+\! a }\!-\!1\right]\frac 12 \!+\!a \!-\!\theta
=\left[\overline{\varepsilon}\!+\!a \!+\!2(\theta\!-\!a )\right]\frac 12
\!+\!a \!-\!\theta=\frac{\overline{\varepsilon}\!+\! a }2>0
$$
[the last inequality follows from (\ref{condi4})$_3$]. By (\ref{condi4})$_5$
and (\ref{defthetatheta1}-\ref{defthetagamma1})
\beq
\ba{l}
\chi_1:=\frac 12\min\left\{ \frac {\mu}4 ,
(\overline{\varepsilon}\!+\!a )\frac{\gamma\!-\!1}2 \!+\!a \!-\!
\theta,\frac {\mu}4 \overline{\varepsilon}^2 \!+\!
\overline{C}\left(\frac{\theta}2\!-\!a \right)
\!+\!\mu(a \!+\!\theta)\right\},\\[6pt]
J(t):=\frac 12\min\left\{ \frac {\:\mu\:}{\varepsilon} ,
(\varepsilon\!+\!a )\frac{\gamma\!-\!1}2 \!+\!a \!-\!\theta, 
\mu\varepsilon^2\!+\!\mu \varepsilon \!+\!
\ddot\varepsilon \!+\!\mu\varepsilon(a \!+\!\theta)
\!-\!(1\!+\!\gamma)\dot C_-\right\},\\[6pt]
B:=\frac 1{\mu}\!+\!\frac 1{\overline{\varepsilon}\!+\!a } \!+\! \frac 1
{\overline{C}},
\ea
\eeq
are positive, and we find
\beq
\dot V\le - (\chi_1\!+\!J)d^2+\frac{1\!+\!\gamma}2
B\int\limits_0^\pi \!\!f^2 dx + \frac{1\!+\!\gamma}2\dot C_+ d^2.
\eeq
We now assume
\beq
B\int\limits_0^\pi \!\!f^2 dx\le \tilde g(t) d^2+
 \tilde g_1(t,d^2)+\tilde g_2(t,d^2)                  \label{hyp4'}
\eeq
where $\tilde g(t),\tilde g_i(t,\eta)$ ($\ i=1,2$ and $ t\ge 0, \ \eta>0$)
denote suitable nonnegative continuous functions.
Without loss of generality we can assume that
$\tilde g_i(t,\eta)$
are non-decreasing in $\eta$; if originally this is not the case,
we just need to replace $\tilde g_i(t,\eta)$ by
$\max\limits_{0\le u\le \eta}\tilde g_i(t,u)$ to fulfill
this condition.

Note that if $f(x,t,U=0)\equiv 0$, as it is the case when $f$ is
obtained as in (\ref{deff}) with $p\equiv 0$, then it is possible to choose
$\tilde g_i$ so that $\tilde g_i(t,0)\equiv 0$.

Summarizing, we have proved
\begin{lemma}
Assume $\varepsilon,\: C,\: \tilde g,\: \tilde g_1(\cdot,\eta),\: \tilde g_2(\cdot,\eta)$
($\eta\!>\!0$) are continuous nonnegative functions of $t\in [0,\infty[$ such that
$\varepsilon,\: C$ fulfill (\ref{condi4}) and $\tilde g,\: \tilde g_1,\: \tilde g_2$ fulfill
(\ref{hyp4'}). Then
\beq
\ba{l}
\chi_0 d^2(t) \le V(t) \le \gamma G(t) d^2(t) \\ [8pt]
\dot V \le - \psi(t) V +g_1\left(t,V\right)+g_2\left(t,V\right),
\ea                                    \label{ineq}
\eeq
where for $i=1,2$
$$
g_i(t,V):=\frac{1\!+\!\gamma}2\,\tilde g_i\!\left(\!t,\frac V{\chi_0}\!\right),
\qquad \psi(t):=b(t)\!-\!g(t),\qquad
b(t):=\frac{\chi_1\!+\!J(t)}{\gamma G(t)},\qquad
g(t):=\frac{1\!+\!\gamma}{2\chi_0}(\dot C_+ \!+\!\tilde g).
$$
\label{lemma1}
\end{lemma}
By the ``Comparison Principle''
(a generalization of Gronwall Lemma,
see e.g. \cite{Yos66}) $V$ is bounded from above
\beq
V(t)\le y(t),            \label{mag1'}
\eeq
by the solution $y(t)$ of the Cauchy problem
\beq
\dot y=-\psi(t)\, y+ g_1(t,y)+g_2(t,y),
\qquad\qquad y(t_0)=y_0\equiv V(t_0)\ge 0;         \label{eqconf'}
\eeq
the latter is equivalent to the integral equation
\beq
y(t)=y_0e^{-\int_{t_0}^t\psi(\tau) d\tau }+
e^{-\int_{t_0}^t\psi(\tau)d\tau } \int_{t_0}^t[g_1\big(\tau,y(\tau)\big)
\!+\!g_2\big(\tau,y(\tau)\big)]
e^{\int_{t_0}^{\tau}\psi(z)dz }d\tau.            \label{inteq}
\eeq
We therefore study the latter. If $\overline{\overline{G}}<\infty$, then
$b(t)>\mbox{const}/\gamma\overline{\overline{G}}=:p>0$, and
the theorems of \cite{DanFio05} apply.
If $G(t)\stackrel{t\!\to\!\infty}{\longrightarrow}\infty$ and
$\overline{\overline{J}}<\infty$, then
$b(t)\stackrel{t\!\to\!\infty}{\longrightarrow}0$, and those theorems
no longer apply.

\begin{lemma}
Assume that $\psi\in C([0,\infty[)$ and $g,\: g_1(\cdot,\eta),\: g_2(\cdot,\eta) $ ($\eta\!>\!0$)
are continuous nonnegative functions of $t\in [0,\infty[$ fulfilling the following properties:
\beq
\ba{l}
\exists\,\bar t\ge 0\qquad\mbox{such that }\qquad\psi(t)>0\quad\:\forall t\ge \bar t;\\[8pt]
\lim\limits_{t\to\infty}\frac{ g_1(t,\eta)}{\psi(t)}=0,
\qquad\forall\eta>0; \\[8pt]
\int\limits^{\infty}_0 g_2(\tau,\eta)d\tau
=:\sigma_2(\eta)<\infty,      \quad\forall\eta>0 .
\ea                             \label{hyp'}
\eeq
Then $\forall\alpha>0$ there exist $ s(\alpha)\ge \bar t$
such that if $0\le y_0\le\alpha$, $t_0\ge  s(\alpha)$, then
the solution $y(t;t_0,y_0)$ of {\rm (\ref{eqconf'})} fulfills
\beq                                                        \label{strbou}
 0\le y(t;t_0,y_0)<\beta(\alpha):=3\alpha, \qquad\qquad t\ge t_0\ge  s(\alpha).
\eeq
If $g_1\equiv g_2\equiv 0$, then $s(\alpha)=0$.
\label{lemma1'}
 \end{lemma}

\bp{}
By (\ref{hyp'})$_1$, if $0\le y_0<\alpha$ then
one finds  for any $t\ge t_0\ge \bar t$
\beq
y_0e^{-\int_{t_0}^t\psi(\tau) d\tau } \le \alpha 
.                    \label{ineq1}
\eeq
On the other hand, by (\ref{hyp'})$_2$
there exists a $s_1(\alpha)\ge \bar t$ such that
$\frac{ g_1(\tau,\beta)}{\psi(\tau)}<\alpha$ for all
$\tau\ge t_0\ge s_1(\alpha)$; then
\bea
&& e^{-\int_{t_0}^t\psi(\tau)d\tau} \int_{t_0}^tg_1\big(\tau,\beta\big)
e^{\int_{t_0}^{\tau}\psi(z)dz }d\tau=
e^{-\int_{t_0}^t\psi(\tau)d\tau} \int_{t_0}^t\psi(\tau)
\frac{g_1\big(\tau,\beta\big)}{\psi(\tau)}
e^{\int_{t_0}^{\tau}\psi(z)dz }d\tau      \label{ineq2}\\
&& \le\alpha  e^{-\int_{t_0}^t\psi(\tau)d\tau}
\int_{t_0}^t\psi(\tau)e^{\int_{t_0}^{\tau}\psi(z)dz }d\tau=
\alpha e^{-\int_{t_0}^t\psi(\tau)d\tau}
\left[e^{\int_{t_0}^{\tau}\psi(z)dz }\right]_{t_0}^t
=\alpha\left[1-e^{-\int_{t_0}^t\psi(\tau)d\tau}  \right]
< \alpha.                          \nonumber
\eea
By (\ref{hyp'})$_3$, there exists a $s_2(\alpha)\ge \bar t$ such that
$\int\limits^{\infty}_{t_0} g_2(\tau,\beta)d\tau<\alpha$ for all
$ t_0\ge s_2(\alpha)$; then
\bea
&& e^{-\int_{t_0}^t\psi(\tau)d\tau} \int_{t_0}^tg_2\big(\tau,\beta\big)
e^{\int_{t_0}^{\tau}\psi(z)dz }d\tau <
e^{-\int_{t_0}^t\psi(\tau)d\tau} \int_{t_0}^tg_2\big(\tau,\beta\big)
e^{\int_{t_0}^{t}\psi(z)dz }d\tau\nn
&& \qquad \qquad = \int_{t_0}^t g_2\big(\tau,\beta\big) d\tau
< \int_{t_0}^\infty g_2\big(\tau,\beta\big) d\tau <\alpha.  \label{ineq3}
\eea

Now let us suppose {\it ad absurdum} that, even if
\ $0\!\le\! y_0\!<\!\alpha$, \ there exists
$t_1> t_0\ge  s(\alpha):=\max\{s_1(\alpha),s_2(\alpha)\}$ such that
\bea
&&0\le y(t;t_0,y_0)<\beta \qquad\qquad\mbox{for } t_0\le t<t_1
                                                            \label{ggg0'}\\
&& y(t_1;t_0,y_0)=\beta.                               \label{ggg'}
\eea
Because of (\ref{ggg0'}) and the monotonicity of $g_i(t,\cdot)$ w.r.t. $\eta$,
for $t\in[t_0,t_1[$
the right-hand side (rhs) of (\ref{inteq}) is bounded from above
by the sum of the rhs's of (\ref{ineq1}-\ref{ineq3}); this implies
\[
y(t_1;t_0,y_0)<\beta,
\]
against the assumption (\ref{ggg'}). Hence, (\ref{strbou}) is proved.
\ep

By the previous lemma and (\ref{mag1'}), (\ref{ineq}), for any
$\alpha>0$ not only the solution $y(t)$ of the
Cauchy problem (\ref{eqconf'}), but also $V(t)$ and $d^2(u,u_t)$,
remain eventually uniformly bounded by $\beta(\alpha)$
if $0\le y_0\le\alpha$.
By the monotonicity of $g_i(t,\eta)$ w.r.t. $\eta$ and the comparison
principle \cite{Yos66} we find that $y(t)$ is also bounded
\beq
y(t)\le z(t), \qquad\qquad t\ge t_0,                   \label{mag2'}
\eeq
by the solution $z(t)$
of the Cauchy problem
\beq
\dot z=-\psi(t)\,z+ g_1(t,\beta)+g_2(t,\beta), \qquad\qquad
z(t_0)=z_0                                     \label{eqconf2'}
\eeq
[which differs from (\ref{eqconf'}) in that the second
argument of $g_i$ is now a fixed constant $\beta>0$], provided that
$z_0=y_0$, and $t_0\ge  s(\alpha)$.
We therefore study the Cauchy problem (\ref{eqconf2'}),
keeping in mind that for our final purposes we will choose
$\beta=\beta(\alpha)=3\alpha$,
$t_0=t_0(\alpha)\ge  s(\alpha)$.

\begin{lemma}
Assume
$\psi\in C([0,\infty[)$ and $g,\: g_1(\cdot,\eta),\: g_2(\cdot,\eta) $ ($\eta\!>\!0$)
are continuous nonnegative functions of $t\in [0,\infty[$ fulfilling  (\ref{hyp'}) and
\beq
\int^\infty_0\psi(t)dt=\infty.                       \label{hyp'2}
\eeq
Then for any $\rho>0$, $t_0\ge \bar t$, $\alpha>0$
there exists $ T(\rho,\alpha,\beta,t_0)>0$
such that, if $0\le z_0<\alpha$, the solution
$z(t;t_0,z_0)$ of {\rm (\ref{eqconf2'})} is bounded as follows:
\beq
0\le z(t;t_0,z_0)<\rho, \qquad\qquad \mbox{if }\quad
t\ge t_0+ T.                        \label{zaza'}
\eeq
\label{lemma2'}
 \end{lemma}

\vskip-.5cm
\bp{}
The solution $z(t)=z(t;t_0,z_0)$ is of the form
\beq
z(t) =
z_0\,e^{- \int\limits^t_{t_0} \psi(\tau)d\tau} + e^{-\int\limits^t_0\psi(\tau)d\tau }
\int\limits_{t_0}^t\left[g_1(\tau,\beta)+
g_2(\tau,\beta)\right]e^{\int\limits_0^{\tau}\psi(\lambda)
d\lambda }d\tau.                                     \label{solu2'}
\eeq
We now consider each of the three terms at the rhs of (\ref{solu2'})
separately. By (\ref{hyp'2}), for all $t_0\ge 0$
$\int^t_{t_0} \psi(\tau)d\tau\stackrel{t\to\infty}{\longrightarrow}
\infty$; hence, there exists a $T_0(\rho, \alpha,t_0)>t_0$
such that
\beq
z_0\,e^{- \int\limits^t_{t_0} \psi(\tau)d\tau}\, <\,\frac{\rho}3,
\qquad\qquad \mbox{if } \quad t\ge T_0,\quad z_0\in[0,\alpha].
\label{magg1'}
\eeq
 By (\ref{hyp'})$_2$, there exist $\sigma_1(\beta)>0$ and
$T_1(\beta,\rho)\ge \bar t$ such that
\beq
\ba{ll}
\frac{ g_1(t,\beta)}{\psi(t)}
 \le \sigma_1\qquad\qquad &\mbox{if}
\:\: t\ge  \bar t, \\[8pt]
\frac{ g_1(t,\beta)}{\psi(t)}
\le \frac 16\,\rho \qquad\qquad &\mbox{if}\:\: t\ge T_1.
\ea                                             \label{chiccoo'} 
\eeq
Moreover, by (\ref{hyp'2}) there exists a $T_2(\beta,\rho)\ge T_1$ such that for
$t\ge T_2$
\beq
e^{-\int\limits^t_0\psi(\tau)d\tau }
\sigma_1(\beta) e^{\int\limits^{T_1}_0\psi(\tau)d\tau } <\frac{\rho}6.       \label{mmmm}
\eeq
Therefore, for $t_0\ge \bar t$ and $t\ge\ T_2\!+\!t_0$,
\bea
&& e^{-\int\limits^t_0\psi(\tau)d\tau} \int\limits_{t_0} ^tg_1(\tau,\beta)
e^{\int\limits_0^{\tau}\psi(\lambda)d\lambda }d\tau \nn
&&\qquad\qquad \le\:\: e^{-\int\limits^t_0\psi(\tau)d\tau}
\int\limits_{\bar t}^{T_1}\psi(\tau) \frac{ g_1(\tau,\beta)}{\psi(\tau)}
e^{\int\limits_0^{\tau}\psi(\lambda)d\lambda }    d\tau
+e^{-\int\limits^t_0\psi(\tau)d\tau}
\int\limits^t_{T_1}\psi(\tau) \frac{ g_1(\tau,\beta)}{\psi(\tau)}
e^{\int\limits_0^{\tau}\psi(\lambda)d\lambda }    d\tau \nn
 &&\qquad\qquad\le\:\: e^{-\int\limits^t_0\psi(\tau)d\tau}
\int\limits_{\bar t}^{T_1}\psi(\tau) \sigma_1(\beta)
e^{\int\limits_0^{\tau}\psi(\lambda)d\lambda }    d\tau
+e^{-\int\limits^t_0\psi(\tau)d\tau}
\int\limits^t_{T_1}\psi(\tau) \frac {\rho}6
e^{\int\limits_0^{\tau}\psi(\lambda)d\lambda }    d\tau \nn
&&\qquad\qquad=\:\: e^{-\int\limits^t_0\psi(\tau)d\tau}\sigma_1(\beta)
\left[ e^{\int\limits_0^{\tau}\psi(\lambda)d\lambda } \right]_{\bar t}^{T_1}
+e^{-\int\limits^t_0\psi(\tau)d\tau}\frac {\rho}6
\left[ e^{\int\limits_0^{\tau}\psi(\lambda)d\lambda} \right]^t_{T_1}\nn
&&\qquad\qquad=\:\: e^{-\int\limits^t_0\psi(\tau)d\tau}\sigma_1(\beta)
\left[ e^{\int\limits_0^{T_1}\psi(\lambda)d\lambda} -
e^{\int\limits_0^{\bar t}\psi(\lambda)d\lambda}  \right]
+e^{-\int\limits^t_0\psi(\tau)d\tau}\frac {\rho}6
\left[ e^{\int\limits_0^{t}\psi(\lambda)d\lambda} -
e^{\int\limits_0^{T_1}\psi(\lambda)d\lambda}\right]\nn
&&\qquad\qquad<\:\: e^{-\int\limits^t_0\psi(\tau)d\tau}\sigma_1(\beta)
 e^{\int\limits_0^{T_1}\psi(\lambda)d\lambda }
+e^{-\int\limits^t_0\psi(\tau)d\tau}\frac {\rho}6
 e^{\int\limits_0^{t}\psi(\lambda)d\lambda}\nn
&& \qquad\qquad< \:\: \frac{\rho}6(1+1)=\frac{\rho}3,\label{dis2z'}
\eea
where we have used (\ref{chiccoo'}),
 the nonnegativity of $g_1(t),\psi(t)$ for $t\ge \bar t$,  and (\ref{mmmm}).
As for the third term at the rhs of (\ref{solu2'}),
by (\ref{hyp'})$_3$ there exists
$T_3(\beta,\rho)> t_0$ such that
\beq
 \int\limits^\infty_{T_3}g_2(\tau,\beta)d\tau
<\frac{\rho}6,                            \label{chiccoo"}
\eeq
and by (\ref{hyp'2}) there exists $T_4(\beta,\rho)> T_3$
such that for $t\ge  T_4$
\beq
e^{-\int\limits^t_{T_3}\psi(\tau)d\tau}
\int\limits^{T_3}_0g_2(\tau,\beta)d\tau
<\frac{\rho}6.     \label{mmmm'}
\eeq
Therefore for $t\ge T_4$
\bea
&& e^{-\int\limits^t_0\psi(\tau)d\tau }
\int\limits_{t_0}^t g_2(\tau,\beta)
e^{\int\limits^\tau_0\psi(\lambda)d\lambda } d\tau
\:\: \le \:\:e^{-\int\limits^t_0\psi(\tau)d\tau } \left[
\int\limits_{\bar t}^{T_3} g_2(\tau,\beta)
e^{\int\limits^\tau_0\psi(\lambda)d\lambda } d\tau+
\int\limits_{T_3}^t g_2(\tau,\beta)
e^{\int\limits^\tau_0\psi(\lambda)d\lambda } d\tau
\right]  \nn
&& \qquad\qquad \:\:<\:\: e^{-\int\limits^t_0\psi(\tau)d\tau } \left[
e^{\int\limits^{T_3}_0\psi(\lambda)d\lambda }
\int\limits_{\bar t}^{T_3} g_2(\tau,\beta)\,  d\tau+
e^{\int\limits^t_0\psi(\lambda)d\lambda }
\int\limits_{T_3}^t g_2(\tau,\beta)\, d\tau
\right]  \nn
&& \qquad\qquad \:\:\le\:\: e^{-\int\limits^t_{T_3}\psi(\tau)d\tau }
\int\limits_0^{T_3} g_2(\tau,\beta)\,  d\tau+
\int\limits_{T_3}^\infty g_2(\tau,\beta)\, d\tau  \nn
&&\label{dis3z'}\qquad\qquad< \:\: \frac{\rho}6 +
\frac{\rho}6=\frac{\rho}3,
\eea
where we have used  the
nonnegativity of $g_2(t),\psi(t)$ for $t\ge \bar t$,
(\ref{chiccoo"}) and (\ref{mmmm'}).

Let $ T(\rho,\alpha,\beta,t_0):=\max\{T_0,T_2,T_4\}$.
Collecting the results (\ref{solu2'}),
(\ref{magg1'}), (\ref{dis2z'}), (\ref{dis3z'}) we find that
the solution $z(t)$ of (\ref{eqconf2'}) fulfills (\ref{zaza'}), as claimed.
\ep

{\bf Remark 3.1} \
By (\ref{hyp'}), (\ref{hyp'2})   the function $\Psi(t):=\int^t_{\bar t}\psi(t)dt$
is nonnegative and increasing for $t\ge \bar t$ and
diverges as $t\to \infty$. If there exists a nonnegative, strictly
increasing function $h:[0,\infty[\mapsto[0,\infty[$ such that $h(0)=0$ and
\beq
\int^{t_0\!+\!\Delta}_{t_0}\psi(t)dt\equiv\Psi(t_0\!+\!\Delta )\!-\!\Psi(t_0)\ge h(\Delta)
\qquad\qquad \forall t_0\ge \bar t, \quad \Delta\ge 0  \label{uniform}
\eeq
(with $h$ not depending on $t_0$), then one can choose in (\ref{magg1'})
$T_0=h^{-1}\left[\log\left(\frac{3\alpha}{\rho}\right)\right]$;
then
{\bf $T$ becomes
independent of $t_0$}. A sufficient condition for (\ref{uniform})
is that there exists a $\psi_0>0$ such that $\psi(t)\ge \psi_0>0$
for all $t\ge t_0$, whence $h(\Delta)=\psi_0\Delta$ and
$T_0=\frac 1{\psi_0}\left[\log\left(\frac{3\alpha}{\rho}\right)\right]$.
However it is not necessary: there are examples that satisfy (\ref{uniform})
but not the latter. A class of such examples is obtained choosing
 $f=r(t)\sin\varphi$, with a function $r(t)$ such that the integral
$\int_0^t r^2(\tau)d\tau$ grows in the average as some power $t^{\chi}$, where
$\chi\le 1$ and in the case $\chi=1$ is smaller than $pt$ for
sufficiently large $t$, but may vanish somewehere; e.g.
we could take $r^2$ a continuous function that vanishes everywhere
except in intervals centered, say, at equally spaced points, where
it takes maxima increasing with some power law $\sim t^{\beta}$,
but keeps the integral bounded, e.g.
\beq
\qquad\quad r^2(t)=r_0^2\:\left\{
\begin{array}{ll}
4n^{\alpha+\beta}(t-n+\frac 1{2n^{\alpha}})\qquad & \qquad\mbox{if}
\qquad t\in[n-\frac 1{2n^{\alpha}},n],\cr
4n^{\beta}- 4n^{\alpha+\beta}(t-n) \qquad &\qquad\mbox{if} \qquad
t\in ]n,n+\frac 1{2n^{\alpha}}],\cr      0\qquad &\qquad\mbox{otherwise,}
\end{array}\right.
\eeq
with $r_0^2<p$, $\alpha\ge 1$, $\beta\in]\alpha-1,\alpha]$
and $n\in\b{N}$ (see  \cite{DanFio05}). The graph
of $(r(t)/r_0)^2$ consists of a sequence of
isosceles triangles enumerated by $n$,
having bases of lenght $1/n^{\alpha}$ and upper vertices with
coordinates $(x,y)=(n,2n^{\beta})$. Their
areas are $A_n=1/n^{\gamma}$, where $\gamma:=\alpha-\beta\in[0,1[$.
Then we can set \ $\tilde g(t):=\pi Br^2(t)$,  \
$\tilde g_1(t)\equiv \tilde g_2(t)\equiv 0$.

We are now in the conditions to prove the following

\begin{theorem}
Assume that $\varepsilon\!\in\! C^2(I)$,
$C\!\in\! C^1(I)$ fulfill (\ref{condi4}),
 the function $f$ of lhs{\rm(\ref{eqd})} belongs to
$ C([0,\pi]\!\times\! I\!\times\!\b{R}^3)$ and is bounded as
in {\rm (\ref{hyp4'})}, where $g(t),\: g_1(t,\eta),\:  g_2(t,\eta)$ \
($t\!\ge\! 0, \ \eta\!>\!0$)
are continuous functions
fulfilling the conditions {\rm (\ref{hyp'}), (\ref{hyp'2}) }. Then the
solutions of the problem {\rm  (\ref{eqd}-\ref{eq2d})} are eventually
bounded (uniformly if $G$ is upper bounded). Moreover, $u\equiv 0$
is eventually quasi-uniform-asymptotically stable in the large. It is
quasi-uniform-asymptotically stable in the large if in addition (\ref{uniform})
is fulfilled.
\label{thm1}
\end{theorem}

\bp{}
Under the assumption $d(u_0,u_1)\le \alpha'$, by
(\ref{ineq}) we find $y_0=V(t_0)\le\alpha$,
where $\alpha:=\alpha'{}^2 \gamma G(t_0)$.
By (\ref{mag1'}) and the application of lemma \ref{lemma1'}
we find that $y(t)$ [and therefore $V(t)$]
is bounded by $\beta(\alpha)$, and again by
(\ref{ineq}) we find
$d(t)\le \beta'(\alpha):=\sqrt{\beta(\alpha)/\chi_0}$
for $t\ge s'(\alpha'):=s(\alpha)$. If $G$ is bounded
the same works with $G(t_0)$ replaced by
$\overline{\overline{G}}$, yielding the (eventual) uniform boundedness, as claimed.
On the other hand, we can now apply
the comparison principle (\ref{mag2'}-\ref{eqconf2'}) and
lemma \ref{lemma2'}: chosen $\rho'>0$, we set
$\rho:= \chi_0\rho'{}^2$, \
$T'(\rho',\alpha',t_0):=T[\rho,\alpha,\beta(\alpha),t_0(\alpha)]$.
As a consequence of (\ref{mag2'}), (\ref{zaza'}),
(\ref{ineq}) we thus find that for $t_0\ge s'(\alpha')$ and
$t\ge t_0\!+\!T'(\rho',\alpha')$,
$$
d^2(t)\le \frac{V(t)}{\chi_0}\le\frac{y(t)}{\chi_0}
\le\frac{z(t,\beta(\alpha'))}{\chi_0}<\frac{\rho}{\chi_0}
=\rho'{}^2,
$$
namely $u\equiv 0$  is eventually
quasi-uniform-asymptotically stable in the large, as claimed.
\ep


{\bf Remark 3.2.}
The two lemmas and the theorem are generalizations respectively of Lemmas 1,2 and
Theorem 1. in \cite{DanFio05}, in that we allow here $t$-dependent $\varepsilon$ and $C$.

We shall use also the following modified version of  Lemma \ref{lemma1'}, where
$s(\alpha):=\bar t$:

\medskip
{\bf Lemma \ref{lemma1'}'} \ \
{\it Assume that $\psi\in C([0,\infty[)$ and $g, g_i$ ($\ i=1,2$)
are continuous nonnegative functions depending only on $t\in [0,\infty[$ and
fulfilling the following properties
$$
\ba{l}
\exists\,\bar t>0\quad\mbox{s. t. }\quad\psi(t)\!>\!0\quad\forall t\ge \bar t,\\[8pt]
\lim\limits_{t\to\infty}\frac{ g_1(t)}{\psi(t)}=0, \qquad\quad
 M_2\!:=\! \int\limits^{\infty}_0 g_2(\tau)d\tau
<\infty.
\ea                             \eqno{(\ref{hyp'}')}
$$
Then \ $\forall\alpha>0$,  \  $t_0\ge  \bar t$, \  if  \ $0\le y_0\le\alpha$  \ then
the solution  \ $y(t;t_0,y_0)$  \ of {\rm (\ref{eqconf'})} fulfills
\beq                                                        \label{strbou"}
 0\le y(t;t_0,y_0)<\tilde\beta(\alpha), \qquad\qquad t\ge t_0, 
\eeq
where
\beq
\tilde\beta:=\alpha \!+\!M_1\!+\!M_2,\qquad\qquad
M_1:=\sup\limits_{t\ge \bar t}\left\{\frac{ g_1(t)}{\psi(t)}\right\}.
\eeq
}

\medskip
\noindent
The lemma is proved using again (\ref{ineq1})
and replacing (\ref{ineq2}),  (\ref{ineq3}) respectively by
\bea
&& e^{-\int_{t_0}^t\psi(\tau)d\tau}\!\!
\int\limits_{t_0}^t\!\!g_1\big(\tau\big) e^{\int_{t_0}^{\tau}\psi(z)dz }d\tau <
M_1 e^{-\int_{t_0}^t\psi(\tau)d\tau}
\!\!\int\limits_{t_0}^t\!\!\psi(\tau)e^{\int_{t_0}^{\tau}\psi(z)dz }d\tau=
M_1\left[1\!-\!e^{-\int_{t_0}^t\psi(\tau)d\tau} \right]<M_1,\nn
&& e^{-\int_{t_0}^t\psi(\tau)d\tau} \int\limits_{t_0}^tg_2\big(\tau\big)
e^{\int_{t_0}^{\tau}\psi(z)dz }d\tau\le \int\limits_{t_0}^t
g_2\big(\tau\big) d\tau < \int_{t_0}^\infty g_2\big(\tau\big) d\tau
<M_2.    \nonumber
\eea

Lemma \ref{lemma2'} holds unmodified.
Theorem \ref{thm1} applies with $s(\alpha)=\bar t$,
i.e. all the properties of the solutions become no more eventual.
In particular this holds in the case $g_1=g_2\equiv 0$ with
$M_1=M_2=0$; the latter situation may occur only if $f(x,t,U=0)\equiv 0$, by
(\ref{hyp4'}).

\begin{theorem}
Assume that $\varepsilon\!\in\! C^2(I)$,
$C\!\in\! C^1(I)$ fulfill (\ref{condi4}),
 the function $f$ of lhs{\rm(\ref{eqd})} belongs to
$ C([0,\pi]\!\times\! I\!\times\!\b{R}^3)$ and is bounded as
in (\ref{hyp4'}), where $g(t),\: g_1(t,\eta),\:  g_2(t,\eta)$ \ ($t\!\ge\! 0, \ \eta\!>\!0$)
are continuous functions fulfilling the conditions  (\ref{hyp'}) and
 in addition $f(x,t,0)\equiv 0$. Then the solution
$u\equiv 0$ is stable [uniformly if $G$ is upper bounded and
(\ref{uniform}) is fulfilled].
\label{thm2}
\end{theorem}

\bp{}
As $f(x,t,0)\equiv 0$ then
$u\equiv 0$ is a solution of (\ref{eq2u}-\ref{eq2d}).
We prove that Lemma \ref{lemma1'} implies its stability.
In fact, we can choose $\beta>0$  as the independent parameter and $\alpha:=\beta/3$
as a dependent one.
Under the assumptions of Lemma \ref{lemma1'}, for any $\beta>0$
$$
0\le y_0<\alpha(\beta),\qquad t_0\ge s''(\beta):=s[\alpha(\beta)] \qquad\qquad
\Rightarrow \qquad\qquad 0\le y(t;t_0,y_0)<\beta,
\qquad\mbox{if }\quad t\ge  t_0.
$$
In particular, choosing $t_0=s''(\beta)$ one finds
$$
y[s''(\beta)]\equiv y_0''<\alpha(\beta)\qquad\qquad
\Rightarrow \qquad\qquad y(t;s''(\beta),y_0'')<\beta,
\qquad\mbox{if }\quad t\ge s''(\beta).
$$

On the other hand, by the continuity of
$y(t;t_0,y_0)$,  there exists a $\delta''(\beta,t_0)\in]0,\alpha(\beta)[$ such that
$$
y_0 <\delta'',\quad t_0\in[\bar t,s''(\beta)],\quad t\in[t_0,s''(\beta)]\qquad \qquad
\Rightarrow \qquad\qquad y(t;t_0,y_0)<\alpha(\beta).
$$
As $\delta''(\beta,t_0)$ is a continuous function
of $t_0$ in the compact domain $[\bar t,s''(\beta)]$,
this inequality holds also with $\delta''$ replaced by the positive function
$\delta(\beta):=\min_{t_0\in[\bar t,s''(\beta)]}\{\delta''(\beta,t_0)\}$.
It holds in particular for $t=s''(\beta)$, hence, setting
$y_0''=y[s''(\beta); t_0, y_0]$,
together with the previous two relations and $\delta(\beta)<\alpha(\beta)$ it implies
\beq
y_0<\delta,\qquad t_0\ge \bar t,\qquad t\ge  t_0 \qquad\qquad
\Rightarrow \qquad\qquad 0\le y(t;t_0,y_0) <\beta.                   \label{inter}
\eeq

Now for any $\sigma>0$ let
$\beta\equiv\beta(\sigma):=\chi_0\sigma^2$, \
 $\delta'(\sigma,t_0):=\sqrt{\delta(\beta)/ \gamma G(t_0)}$.
By (\ref{ineq}) we find that  $d(u_0,u_1)\le \delta'$
implies $y_0<\delta$ and therefore rhs(\ref{inter});
by (\ref{mag1'}) and again by
(\ref{ineq}) we find
\beq
d(t)\le \sigma,\qquad\qquad t\ge t_0\ge \bar t.
\eeq
This amounts to the
stability of the null solution. Finally $\overline{\overline{G}}<\infty$
and (\ref{uniform}) imply that $\delta'',\delta$ are
independent of $t_0$, so is $\delta':=\sqrt{\delta(\beta)/ \gamma
\overline{\overline{G}}}$, and the null solution is uniformly stable.
\ep

\sect{Total stability}
\label{totalstability}

Consider the special case that one can choose $g_1,g_2$ so that
\beq
g_1(\cdot,0)\equiv 0, \qquad\qquad g_2(t,y)=g_{21}(t,y)+ g_{22}(t),
\qquad\mbox{with}\quad g_{21}(\cdot,0)\equiv 0    \label{special}
\eeq
($g_{22}$ is necessarily nonnegative).
Problem (\ref{eqconf'}) becomes
\beq
\dot y=-\psi(t)\, y+ g_1(t,y)+g_{21}(t,y)+ g_{22}(t), \qquad\qquad
y(t_0)=y_0\equiv V(t_0)\ge 0.        \label{eqconf"}
\eeq
We shall denote its solution as $y(t;t_0,y_0; g_{22})$ when we wish to emphasize the
dependence on $g_{22}$.

Assume that $\psi, g_1(\cdot,\eta)$ and
$\hat g_2(t,\eta)\!=\!g_{21}(t,\eta)\!+\!\hat g_{22}(t)$
fulfill (\ref{special}) and the conditions of Lemma \ref{lemma1'};
for any $\alpha>0$
we apply the lemma and denote as $\hat s(\alpha)$ the corresponding
value of $s(\alpha)$.

First, we  define the set $\widehat{G}$ of admissible perturbations related to
$\hat g_{22}$ by setting
\beq
\widehat{G}:=
\left\{ r\in C([\bar t,\infty[)  \:\:\quad \left\vert  \:\:\quad r\ge 0, \quad
\int^{\infty}_{t_0}\!\!\! r(\tau)\, d\tau\le \int^{\infty}_{t_0}\!\!\! \hat g_{22} \,d\tau
\quad \forall t_0\ge \bar t
\right.\right\}                      \label{G2}
\eeq
and   note that for all $g_{22} \in \widehat{G}$
inequality (\ref{ineq3}), and therefore also the claim (\ref{strbou}),
hold  {\it again} for $t_0\ge \hat s(\alpha)$, because
\beq
\int^{\infty}_{t_0}g_2(\tau,\beta) d\tau\le \int^{\infty}_{t_0}\hat g_2(\tau,\beta)  d\tau  \qquad\qquad\forall t_0\ge \bar t , \quad \beta>0  .                 \label{bla}
\eeq
Choosing $\beta\!>\!0$ and setting
$\alpha(\beta)\!:=\!\beta/3$,
from Lemma \ref{lemma1'} we find that for any
$t_0\ge\hat s(\beta)\!:=\!\hat s[\alpha(\beta)]=\hat s(\beta/3)$
\beq                                                        \label{strbou'}
0\le y_0<\alpha,\qquad g_{22} \in \widehat{G} \qquad \qquad \Rightarrow\qquad \qquad
 0\le y(t;t_0,y_0; g_{22} )<\beta, \qquad\quad t\ge t_0\ge\hat s(\beta).
\eeq
Second, we ask what we can say if  $\bar t\le t_0\le\hat  s(\beta)$.
Eq. (\ref{eqconf"}) can be seen as a perturbation of the equation
\beq
\dot y^*=-\psi(t)\, y^*+ g_1(t,y^*)+g_{21}(t,y^*),        \label{eqconf"'}
\eeq
which admits the solution $y^*(t)\equiv 0$; we show that
to $\alpha(\beta) <\beta$ and $\bar t\le \hat s(\beta)$
there corresponds a $\hat \delta(\beta)< \alpha(\beta) $ such that
\beq
\ba{c}
\bar t\!\le\! t_0\!<\!\hat  s(\beta) , \qquad\quad 0\!\le\! y_0\!<\!\hat \delta(\beta), \qquad\quad
g_{22} \!\in\! \widehat{G},\qquad\mbox{and}\qquad
\left\{ \ba{l}
\int\limits^{\infty }_{\bar t}\!\! g_{22}(\!\tau\!) d\tau <\hat  \delta(\!\beta\!)\\[10pt]
\mbox{or}\quad\sup\limits_{\tau\ge \bar t}g_{22}(\!\tau\!) <\hat  \delta(\!\beta\!)
\ea\right.
  \qquad \quad \Rightarrow \\[14pt]
 0 \le y(t;t_0,y_0; g_{22} ) <\alpha \qquad\quad
\forall t\!\in\! [t_0,\hat  s(\beta) ],\qquad\qquad\qquad\qquad \\[10pt]
\mbox{ in particular }\qquad
y_0^\beta:=y[\hat s(\beta) ;t_0,y_0; g_{22} ] <\alpha. \qquad\qquad \qquad\qquad
\ea              \label{cont1}
\eeq
In fact, by  Theorem 5.2 in \cite{Yos66} on
the continuous dependence of the solution of (\ref{eqconf"'}) both on the initial data and on the
perturbation term, to $\alpha(\beta) <\beta$ and $\bar t\le \hat s(\beta)$
there corresponds a $\hat \delta_1(\beta)< \alpha(\beta) $ such that
$$
\ba{c}
\bar t\!\le\! t_0\!<\!\hat  s(\beta) , \qquad 0\!\le\! y_0\!<\!\hat \delta_1(\beta), \qquad
g_{22} \!\in\! \widehat{G},\qquad
\int\limits^{\hat s(\beta) }_{t_0}\!\!\! g_{22}(\tau) \, d\tau <\hat  \delta_1(\beta),   \qquad\qquad
  \qquad \quad \Rightarrow \\[12pt]
 y(t;t_0,y_0; g_{22} )\equiv | y(t;t_0,y_0; g_{22} )\!-\!0| <\alpha \qquad
\forall t\!\in\! [t_0,\hat  s(\beta) ].
\ea
$$
First, if we choose $g_{22} \!\in\! \widehat{G}$ such that
$\int\limits^{\infty }_{\bar t}\!\! g_{22}(\!\tau\!) d\tau <\hat  \delta_1(\!\beta\!)$
then of course the previous inequality is fulfilled. Second, if we choose $g_{22} \!\in\! \widehat{G}$
such that $g_{22}(\!\tau\!)< \delta_2(\!\beta\!)\!:=\!\hat\delta_1(\!\beta\!)/\hat s(\beta)$ for all
$\tau\!\ge\! \bar t$ then
$$
\int\limits^{\hat s(\beta) }_{t_0}\!\!\! g_{22}(\tau) \, d\tau < \delta_2(\!\beta\!)[\hat s(\beta)\!-\!t_0]\le
\hat  \delta_1(\beta)
$$
and again the same inequality is fulfilled. These two arguments lead to (\ref{cont1})
provided we set
$\hat\delta(\!\beta\!):=\min\{\hat\delta_1(\!\beta\!),\delta_2(\!\beta\!)\}$.
Using  (\ref{strbou'})
- after the replacement \ $(t_0,y_0)\mapsto\big( \hat s(\beta) , y_0^\beta \big)$ - \ it follows that the assumptions at the lhs(\ref{cont1}) imply more generally
$$
0\le y(t;t_0,y_0; g_{22} )=y[t;s(\beta),y_0^\beta ; g_{22} ]<\beta,
 \qquad\forall t\ge \hat s(\beta) . 
$$
This relation and (\ref{strbou'}) show that we have proved

\begin{lemma}
Assume that \ $\psi\in C(I)$, \
$ g_1(\cdot,\eta),\: g_2(\cdot,\eta) \in C(I)$ \ ($\eta\!>\!0$)
are nonnegative functions fulfilling  (\ref{hyp'}),
(\ref{special}). Then, given a class of perturbations $\widehat{G}$,
for any \ $ \beta>0$  \ there exists a \
$ \hat \delta(\beta)\!\in\, ]0,\beta[$ \
such that the solution of (\ref{eqconf"}) fulfills
\beq
\ba{c}
y_0\in ]0,\hat \delta[,  \qquad  t_0\ge \bar t,\quad  g_{22} \!\in\!\widehat{G},
\qquad\int\limits^{\infty }_{\bar t}\!\! g_{22}(\tau)  \, d\tau <\hat  \delta\quad
{\rm or}\quad\sup\limits_{\tau\ge \bar t}g_{22}(\!\tau\!) <\hat  \delta(\!\beta\!)
\qquad \quad\Rightarrow\\[12pt]
\qquad \quad 0\le y(t;t_0,y_0; g_{22} ) <\beta  \qquad\forall t\ge t_0.
\ea\label{ineqlemma4}
\eeq
\label{lemma4}
\end{lemma}


We also introduce a  notion of {\it total stability},
of the solution $u(x,t)\equiv 0$. Suppose  that \
$f(x,t,U)=\bar f(x,t,U)\!+\!j(x,t)$  \ with $\bar f(x,t,0)\equiv 0$, \  so that
problem (\ref{eqd}+\ref{eq2d})
admits the solution  \ $ u(x,t)\equiv 0$ when $j\equiv 0$.

\bdefi \label{def-4} \rm $u(x,t)\equiv 0$  is {\it totally stable} if 
for any \  $\sigma>0$ \ there exist \  $\delta(\sigma), \nu(\sigma) > 0$ \  such that
$$
d(u_0,u_1)<\delta,\qquad\mbox{and}
\qquad  \left\{ \ba{l}
\displaystyle\int_{t_0}^\infty\!\!\! dt\int_0^\pi \!\!j^2(x,t) dx < \nu \\[12pt]
\mbox{or}\quad\sup\limits_{t\ge t_0}\displaystyle\int_0^\pi \!\!j^2(x,t) dx < \nu
\ea\right.\qquad \Rightarrow\qquad
d(u,u_t)<\sigma\:\quad\forall t\ge t_0.
$$
\edefi

We can now formulate the following total stability theorem:

\begin{theorem}
Assume that $\varepsilon\!\in\! C^2(I)$, $C\!\in\! C^1(I)$ fulfill (\ref{condi4}),  the function $f$ of lhs{\rm(\ref{eqd})}  is continuous and has the form
$f(x,t,U)=\bar f(x,t,U)\!+\!j(x,t)$,  \ with $\bar f(x,t,0)\equiv 0$ and
\beq
B\int_0^\pi \!\! 2\bar f^2 dx\le g(t) d^2+
 g_1(t,d^2)+g_{21}(t,d^2) ,                \label{hyp4"}
\eeq
 where  (with notation as in Lemma \ref{lemma1}) \
$g(t),\: g_1(t,\eta),\:  g_{21}(t,\eta)$ \
are continuous functions fulfilling the conditions {\rm (\ref{hyp'})},
$g_1(t,0)\equiv 0,\:  g_{21}(t,0)\equiv 0$, and
\beq
 \int_{\bar t}^\infty\!\!\! dt\int_0^\pi \!\!j^2(x,t) dx < \infty.     \label{hyp5}
\eeq
Then  $u\equiv 0$ is totally stable.
\label{thm3}
\end{theorem}

\bp{}
Eq. (\ref{hyp4"}) and $f^2\le 2(\bar f^2\!+\!j^2)$ imply
$$
B \int_0^\pi\!\!\! f^2\, dx \le B\int_0^\pi\!\!\!
2(\bar f^2\!+\!j^2)dx \le  g(t) d^2+
 g_1(t,d^2)+ g_{21}(t,d^2)+ \hat g_{22}(t),
$$
where \ $\hat g_{22}(t):=2B\int_0^\pi \!\!j^2(x,t) dx$; hence
the assumptions of Lemma \ref{lemma4} are fulfilled.
Therefore, recalling (\ref{LBV}), for any $\sigma>0$ we choose
$\beta\equiv\beta(\sigma):=\chi_0\sigma^2$ \  and \
 $\delta'(\sigma,t_0):=\sqrt{\hat\delta[\beta(\sigma)]/ \gamma G(t_0)}$,
where $\hat\delta(\beta)$ is determined by the Lemma.
By (\ref{ineq}), (\ref{eqconf'})$_2$ we find that  $d(u_0,u_1)\le \delta'$
implies $y_0<\hat\delta$ and, by the Lemma,
rhs(\ref{ineqlemma4});
by (\ref{mag1'}) and again by
(\ref{ineq}) we find
$$
d(t)\le \sigma,\qquad\qquad t\ge t_0\ge \bar t.
$$
\ep
{\bf Remark 4.1} \ It is easy to realize that if $j$ is bounded w.r.t. $u$,
namely there exists a constant $K\!>\!0$ and a
continuous function $j_0(x,t)$ such that $|j(x,t,U)|\le K\, j_0(x,t)$,
then the previous theorem applies as well.

\subsection{Application to the problem with initial and boundary perturbations}
\label{appl}

Theorem \ref{thm3} can be applied to discuss the following two special situations.

\subsubsection{Application to a non-analytic forcing term}

\ Let us consider the special case that the function $h$ of
(\ref{eq}-\ref{eqp}) be of the form
\beq
h(x,t,\Phi)\equiv h_0(t) |\varphi(x,t)|^\omega \varphi(x,t), \qquad \omega\in\b{R}^+.
\eeq
By (\ref{deff}) the corresponding $f$ is given by \
$f(x,t,U) =  h_0|u\!+\!\varphi\!+\! p|^\omega (u\!+\!\varphi\!+\! p)
-  h_0|\varphi|^\omega  \varphi- Lp+k$, \ whence
\bea
f^2 &\le & \left[h_0(|u|\!+\!|\varphi\!+\! p|)^{\omega+1}
-  h_0|\varphi|^\omega  \varphi\!-\! Lp\!+\!k\right]^2
 \le \left[h_02^\omega\!\left(|u|^{\omega+1}\!+\!|\varphi\!+\! p|^{\omega+1}\right)
-   h_0|\varphi|^\omega  \varphi\!-\! Lp\!+\!k\right]^2 \nn
&=& \left[h_0  2^\omega |u|^{\omega+1}\!+\!(h_0 2^\omega|\varphi\!+\! p|^{\omega+1}
-  h_0|\varphi|^\omega  \varphi\!-\! Lp\!+\!k)\right]^2
\le 2^{2\omega+1}h_0^2 |u|^{2\omega+2}\!+\!2\left[h_0 2^\omega|\varphi\!+\! p|^{\omega+1} -  h_0|\varphi|^\omega  \varphi\!-\! Lp\!+\!k)\right]^2.\nonumber
\eea
In the second and third inequality we have used the known one
\beq
\left[\sum_{k=1}^n a_k\right]^s\le n^{s-1} \sum_{k=1}^n (a_k)^s
\qquad \qquad\mbox{ if }s>1, \quad n\in\b{N}, \quad a_k\ge 0.    \label{miranda}
\eeq
But applying (\ref{3ineq}) to $u$ one finds  $|u(x,t)|\le d(u,u_t)$,
$|u(x,t)|\le \frac{\pi^{3/2}}{\varepsilon(t)} d(u,u_t)(t)$ for all
$x\in[0,\pi]$, whence
\bea
\int\limits_0^\pi\!\! f^2\, dx&\le  & \int\limits_0^\pi\!\!  2^{2\omega+1}h_0^2 |u|^{2\omega+2}\, dx\!+\!2\int\limits_0^\pi\!\! \left[h_0\, 2^\omega|\varphi\!+\! p|^{\omega+1}
-  h_0^2|\varphi|^\omega  \varphi\!-\! Lp\!+\!k)\right]^2dx\nn
&\le  &2\int\limits_0^\pi\!\! \left[h_0 2^\omega|\varphi\!+\! p|^{\omega+1}
-  h_0^2|\varphi|^\omega  \varphi\!-\! Lp\!+\!k)\right]^2dx + h_0^2\,  2^{2\omega+1}  d^{2\omega+2}
\times \left\{\ba{l} \pi\\  \frac{\pi^{3\omega+4}}{\varepsilon^{2\omega+2}} \ea\right. .
\nonumber
\eea
If $\varphi\equiv 0$, then  (\ref{hyp4'}) holds with
$$
\tilde g\equiv0,\qquad \frac {\tilde g_1(d,t)}{B}:=
\frac 12 h_0^2(t)\, (2d)^{2\omega+2}
\times \left\{\!\!\ba{l} \pi\\  \frac{\pi^{3\omega+4}}{\varepsilon^{2\omega+2}} \ea\right.\!\! ,\qquad\frac  {\tilde g_2(t)}{2B} =\frac
{\tilde g_{22}(t)}{2B}:= \!\int\limits_0^\pi\!\! \left[h_0 2^\omega|p|^{\omega+1}
\!-\! Lp\!+\!k\right]^2dx,
$$
namely (\ref{special}) is fulfilled (with $g_{21}\equiv 0$); if the conditions (\ref{hyp'}) are fulfilled, then we can apply Lemma \ref{lemma4} and  Theorem \ref{thm3}.

If $\varphi\neq 0$, then (\ref{hyp4'}) holds with the same $\tilde g,\tilde g_1 $ and
$\tilde g_2(t)=2B\int\limits_0^\pi\!\! \left[h_0 2^\omega|\varphi\!+\! p|^{\omega+1}
-  h_0^2|\varphi|^\omega  \varphi\!-\! Lp\!+\!k)\right]^2dx$;
if the conditions (\ref{hyp'}') are fulfilled, then
we can apply Lemma \ref{lemma1'}' and
Theorem \ref{thm1}, with the conclusions of the latter holding {\it non-eventually}.

\subsubsection{Application to Lipschitz forcing terms}

We  can apply  Lemma \ref{lemma4} also to the case that $h(x,t,W)$ be a Lipschitz
function w.r.t. the $W$ variables with a `constant' (i.e. a maximal Lipschitz function)
$h_0$ depending only on $t$:
\beq
|h(x,t,W\!+\!\Phi)-h(x,t,\Phi)|\le h_0(t )\left[|w|\!+\! |w_x|\!+\!
|w_t|\right];
\eeq
in fact, by (\ref{deff})
\bea
|f(x,t,U) | &\!=\!& | h(x,t,U\!+\! P\!+\!\Phi)\!-\!h(x,t,\Phi)\!-\! (Lp)(x,t)\!+\!k (x,t) |
\nn & \le &
h_0\!\left[|u\!+\!p|\!+\! |u_x\!+\!p_x|
\!+\!|u_t\!+\!p_t|\right]+| Lp\!-\!k| \nn
& \le & h_0\!\left[|u|\!+\!|u_x|\!+\!
|u_t|\!+\! |p|\!+\! |p_x| \!+\! 
|p_t|\right] +| Lp\!-\!k|,\nonumber
\eea
what implies
\bea
f^2 & \le &  7\left\{ h_0^2\!\left[u^2\!+\!u_x^2\!+\! 
u_t^2\!+\!p^2\!+\! p_x^2 \!+\! 
p_t^2\right]+ (Lp\!-\!k)^2\right\}\qquad\Rightarrow \nn[8pt]
\int_0^\pi\!\! f^2\, dx &\le  & \int_0^\pi
7\left\{ h_0^2\!\left[u^2\!+\!u_x^2\!+\!u_t^2\!+\!
p^2\!+\!p_x^2 \!+\! 
p_t^2\right]+3p_{tt}^2+3(a p_t)^2+3 k^2\right\}dx\nn[8pt]
 & \le & \tilde g(t) d^2 +g_{22}(t),\\[8pt]
\tilde g(t) &:=& 7 h_0^2
\qquad\qquad g_{22}(t):= \int_0^\pi\!\!
7\left\{ h_0^2\!\left[p^2\!+\!
p_x^2 \!+\! 
p_t^2\right]\!+\!3 p_{tt}^2\!+\!3(a p_t)^2\!+\!3 k^2\right\}dx;
\nonumber
\eea
here we have used again (\ref{miranda}) in the first line
and the relation $Lp= p_{tt}+a p_t$ in the second line. This shows that
(\ref{special}) is fulfilled with $g_1\equiv 0$, $g_{21}\equiv 0$;
if the conditions (\ref{hyp'}) are fulfilled, then we can apply Lemma \ref{lemma4}
 and Theorem \ref{thm3}. 
As an example we may choose  $h(x,t,W)=h_0(t)\sin w$, what would
make (\ref{eq}) a modified sine-Gordon equation.

\sect{Examples}
\label{examples}

\begin{figure}[ht]
\begin{center}
\includegraphics[width=5.5cm]{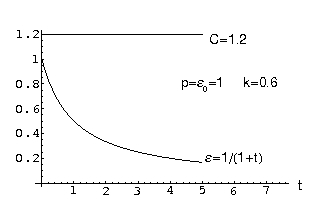}\hfill
\includegraphics[width=5.5cm]{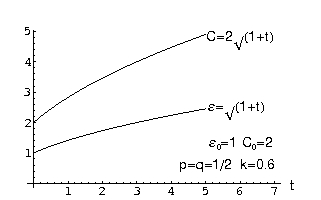}
\end{center}
\caption{$t$-dependences for $\varepsilon,C$ in Examples 1,2}
\label{Example1}       
\end{figure}

Many $\varepsilon(t), C(t),a , f$ fulfill  (\ref{condi4}), (\ref{hyp4'}),
and (\ref{hyp'}), but not the hypotheses of older theorems:

\medskip
{\bf Example 1:} We assume that $a \!>\!0$, \
$\varepsilon(t)=\varepsilon_0(1\!+\!t)^{-\nu }$, with
$\nu \!\ge\!0$, $\varepsilon_0\!\ge\!0$, $C\!>\!0$,
are constant (see
Fig. \ref{Example1}-left, where we have chosen $p=\varepsilon_0=1$, $C=1.2$)
and condition (\ref{hyp4'}) is fulfilled with
$$
\tilde g_i(t)\equiv 0,\qquad
\qquad\tilde g(t)\le g_0,\qquad\mbox{with some constant }
g_0<\frac{2\chi_0\chi_1}{\gamma(\gamma\!+\!1)(C\!+\!1)}.
$$
Then conditions (\ref{condi4}) are fulfilled  defining
$\mu\!=\!\min\{ C/2,(\nu\!+\!1)(C/2\varepsilon_0)^{1/(1\!+\!\nu)}\}$.
Note that $\varepsilon,\dot\varepsilon,\ddot\varepsilon\!\to\! 0$ as
$t\!\to\! \infty$. We find \
$G(t)\!=\!C\!+\!\nu \varepsilon_0[1\!+\!t]^{-\!\nu \!-\!1}\!+\!1\to C\!+\!1$ \ and,
$$
G(t)\to C\!+\!1,\quad J(t)\to 0,\quad b(t)\to \frac{\chi_1}{\gamma(C\!\!+\!1)},\quad
\psi(t)\to \frac{\chi_1}{\gamma(C\!\!+\!1)}\!-\!\frac{g_0(\gamma\!+\!1)}{2\chi_0}\!>\!0,
$$
as \ $t\to\infty$, \ so that  (\ref{hyp'}) and (\ref{uniform}) are fulfilled.
Theorem \ref{thm1} applies: the solutions of  (\ref{eq}-\ref{eq2}) are eventually uniformly bounded, and $u(x,t)\!\equiv\!0$  is
eventually quasi-uniform-asymptotically stable in the large.
If in addition $f(x,t,0)\equiv 0$, then Theorem \ref{thm2} applies and the solution
$u\equiv 0$ is uniformly stable.

\medskip
{\bf Example 2:} We assume that $\varepsilon(t)=\varepsilon_0(1\!+\!t)^\nu$, \
$C(t)=C_0(1\!+\!t)^q$,   \
condition (\ref{hyp4'}) is fulfilled with
$\tilde g_i(t)\equiv 0$,  \ $\tilde g(t)\!\le\! K(1\!+\!t)^r$,
where $K,p,q,r,C_0,\varepsilon_0,a $ are some constants
such that \
$$
K\!>\! 0, \quad 1\!>\!q\!\ge\! p\!>\! 0, \quad 2p\!-\!q\!>\!r\!\ge\!0, \quad
\varepsilon_0\!\ge\! 0, \quad C_0\!>\! p\varepsilon_0, \quad a \!>\!-\varepsilon_0;
$$
this is illustrated in Fig. \ref{Example1}-right, where more specifically we have
chosen $\varepsilon_0=1$, $C_0=2$, $p=q=1/2$.
Then conditions (\ref{condi4}) are fulfilled  defining
$\mu=(C_0\!-\!\nu\varepsilon_0)/(1\!+\!\varepsilon_0)$.
Note that \ $C(t),\varepsilon(t)\!\to\!\infty$, \ but  \
$\dot \varepsilon(t),\dot C(t)\!\to\!0$ \  as \ $t\!\to\!\infty$.
$G(t)\sim t^q$, $b(t)\sim t^{2\nu\!-\!q}$,  \ whence
\ $\psi(t)\to \infty$  as \ $t\to\infty$: therefore
  (\ref{hyp'}) and (\ref{uniform}) are fulfilled.
Theorem \ref{thm1} applies: the solutions of  (\ref{eq}-\ref{eq2}) are eventually uniformly bounded, and $u(x,t)\!\equiv\!0$  is
eventually quasi-uniform-asymptotically stable in the large.
If in addition $f(x,t,0)\equiv 0$, then Theorem \ref{thm2} applies and the solution
$u\equiv 0$ is uniformly stable.

\begin{figure}[ht]
\includegraphics[width=5.5cm]{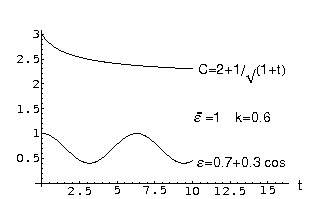}
\caption{$t$-dependences for $\varepsilon,C$ in Examples 3}
\label{Example3}
\end{figure}

{\bf Example 3:} \ \   We assume that  \ $a \!>\!0$;  \ that \
$\varepsilon,\dot \varepsilon,\ddot\varepsilon$  \ are bounded
(in particular may be {\it periodic}); that \
$C(t)\!=\!C_0\!+\! C_1(1\!+\!t)^{-q}$ \ with \ $C_1\!>\!0$, $q\!\ge\!0$,
$C_0\!>\!\max\left\{0,\overline{\overline{\dot\varepsilon}}\right\}$;
that (\ref{hyp4'})
is fulfilled with
$$
\tilde g_i(t)\equiv 0,\qquad
\quad\tilde g(t)\le g_0,\quad\mbox{with some constant }
g_0<\frac{2\chi_0\chi_1}{\gamma(\gamma\!+\!1)(C_0\!+\! C_1\!+\!1
\!-\!\overline{\dot\varepsilon}/2)}.
$$
This is illustrated in Fig. \ref{Example3}, where more specifically we have
chosen \ $\varepsilon(t)=0.7+0.3\cos t$, $C_0\!=\!2$, $C_1\!=\!1$, $q\!=\!-1/2$.
We find $G(t)\le
C_0\!+\!C_1\!-\!\overline{\dot\varepsilon}\!+\!1
\!<\!\infty$.
Conditions (\ref{condi4}) are fulfilled with
$\mu\!=\!(C_0\!-\!\overline{\overline{\dot\varepsilon}})/
(1\!+\!\overline{\overline{\varepsilon}})$.
We find $\dot C_+=0$, $G(t)\le
C_0\!+\!C_1\!-\!\overline{\dot\varepsilon}/2\!+\!1
\!<\!\infty$,
$\psi(t)\ge \frac{\chi_1}{\gamma(C\!+\!1)}\!-\!
\frac{g_0(\gamma\!+\!1)}{2\chi_0}>\!0$:
 (\ref{hyp'}) and (\ref{uniform}) are fulfilled.
Theorem \ref{thm1} applies: the solutions of  (\ref{eq}-\ref{eq2}) are eventually uniformly bounded, and $u(x,t)\!\equiv\!0$  is
eventually quasi-uniform-asymptotically stable in the large.
If in addition $f(x,t,0)\equiv 0$, then Theorem \ref{thm2} applies and the solution
$u\equiv 0$ is stable. 

\end{document}